\pdfoutput=1
\documentclass[journal]{IEEEtran}

\usepackage[utf8]{inputenc}

\usepackage{cite}

\usepackage{booktabs}
\usepackage[pdftex]{graphicx}

\usepackage{tikz}
\usepackage{pgfplots}
\usepackage{eso-pic}
\pgfplotsset{compat=1.13}

\usepackage{etoolbox}

\usepgfplotslibrary{external}
\usepgfplotslibrary{colormaps}
\usepgfplotslibrary{units}
\usepgfplotslibrary{groupplots}
\usepgfplotslibrary{statistics}
\usetikzlibrary{chains}
\usetikzlibrary{scopes}
\usetikzlibrary{patterns}
\pgfplotsset{
    colormap={parula}{
        rgb255=(53,42,135)
        rgb255=(15,92,221)
        rgb255=(18,125,216)
        rgb255=(7,156,207)
        rgb255=(21,177,180)
        rgb255=(89,189,140)
        rgb255=(165,190,107)
        rgb255=(225,185,82)
        rgb255=(252,206,46)
        rgb255=(249,251,14)
    },
    colormap={hotinv}{
        [1cm]rgb255(0cm)=(255,255,255)
        rgb255(2cm)=(255,255,0)
        rgb255(5cm)=(255,0,0)
        rgb255(8cm)=(0,0,0)
    },
}

\pgfplotsset{%
    every x tick label/.append style={/pgf/number format/1000 sep=},
    unit markings=slash space,
    mymarks/.style={mark=*, mark size=1pt, mark options={solid}},
    shape0.10/.style={dashdotted},
    shape0.25/.style={line width=0.75pt, densely dotted},
    shape0.50/.style={dashed},
    shape1.00/.style={black},
    comp1.0000/.style={solid},
    comp0.5000/.style={dashed},
    comp0.2500/.style={line width=0.75pt, densely dotted},
    comp0.0010/.style={dashdotted},
    comp0.1000/.style={dashdotted},
    snr5/.style={red},
    snr20/.style={blue},
    Wiener/.style={black, dashdotted, line width=1pt},
    LSA/.style={black, densely dotted, line width=1pt},
    STSA/.style={black, dashed},
    MOSIE/.style={orange, line width=1pt},
    MOSIESTSA/.style={white!25!blue, dashdotted, line width=1pt},
    specgram_axis/.style={
        axis on top,
        colorbar,
        colormap name={hotinv},
        ymin=0,
        ymax=4,
        xmin=0.5,
        xmax=4.5,
        point meta min=-40,
        point meta max=20,
        xlabel={$t~/~\text{s}$},
        ylabel={$f~/~\text{kHz}$},
        colorbar style={ylabel={$|\sNoisyFreqI|^2~/~\text{dB}$}},
    },
    priorsnr_axis/.style={
        specgram_axis,
        colormap name={parula},
        point meta min=-20,
        point meta max=20,
        colorbar style={ylabel={$\sPriorSNRIEst~/~\text{dB}$}},
    },
    legend image code/.code={
        \draw[mark repeat=3,mark phase=2]
        plot coordinates {
            (0cm,0cm)
            (0.2cm,0cm)        %% default is (0.3cm,0cm)
            (0.4cm,0cm)         %% default is (0.6cm,0cm)
        };%
    },
}
\input{colorconverter}

\newenvironment{customlegend}[1][]{%
    \begingroup
    \csname pgfplots@init@cleared@structures\endcsname
    \pgfplotsset{#1}%
}{%
    \csname pgfplots@createlegend\endcsname
    \endgroup
}%

\def\addlegendimage{\csname pgfplots@addlegendimage\endcsname}

\usepackage{amsmath}
\usepackage{amssymb}

\usepackage{algorithm}
\usepackage{algorithmic}

\usepackage{array}

\ifCLASSOPTIONcompsoc
  \usepackage[caption=false,font=normalsize,labelfont=sf,textfont=sf]{subfig}
\else
  \usepackage[caption=false,font=footnotesize]{subfig}
\fi

\usepackage{url}

\usepackage{acro}

\usepackage{microtype}

\DeclareAcronym{CDF}{short=CDF, long=cumulative distribution function}
\DeclareAcronym{DFT}{short=DFT, long=discrete Fourier transform}
\DeclareAcronym{PSD}{short=PSD, long=power spectral density}
\DeclareAcronym{SNR}{short=SNR, long=signal-to-noise ratio}
\DeclareAcronym{MSE}{short=MSE, long=mean-squared error}
\DeclareAcronym{MMSE}{short=MMSE, long=minimum mean-squared error}
\DeclareAcronym{SPP}{short=SPP, long=speech presence probability}
\DeclareAcronym{IDFT}{short=IDFT, long=inverse discrete Fourier transform}
\DeclareAcronym{STFT}{short=STFT, long=short-time Fourier transform}
\DeclareAcronym{VTS}{short=VTS, long=vector Taylor series}
\DeclareAcronym{LSA}{short=LSA, long=log-spectral amplitude estimator}
\DeclareAcronym{STSA}{short=STSA, long=short-term spectral amplitude estimator}
\DeclareAcronym{TCS}{short=TCS, long=temporal cepstrum smoothing}
\DeclareAcronym{PESQ}{short=PESQ, long=Perceptual Evaluation of Speech Quality}
\DeclareAcronym{PDF}{short=PDF, long=probability density function}
\DeclareAcronym{HMM}{short=HMM, long=hidden Markov model}
\DeclareAcronym{GMM}{short=GMM, long=Gaussian mixture model}
\DeclareAcronym{MoG}{short=MoG, long=mixture of Gaussians, long-plural-form=mixtures of Gaussians}
\DeclareAcronym{NMF}{short=NMF, long=nonnegative matrix factorization}
\DeclareAcronym{DNN}{short=DNN, long=deep neural network}
\DeclareAcronym{MOSIE}{short=MOSIE, long=(M)MSE estimation with (o)ptimizable (s)peech (m)odel and (i)nhomogeneous (e)rror criterion}
\DeclareAcronym{BECOCO}{short=BECOCO, long=phase-(b)lind (e)stimator of (co)mplex (co)efficients}
\DeclareAcronym{MFCC}{short=MFCC, long=Mel-frequency cepstral coefficient}
\DeclareAcronym{CMVN}{short=CMVN, long=cepstral mean and variance normalization}
\DeclareAcronym{ReLU}{short=ReLU, long=rectified linear unit}
\DeclareAcronym{IS}{short=IS, long=Itakura-Saito}
\DeclareAcronym{STOI}{short=STOI, long=short-time objective intelligibility}
\DeclareAcronym{ML}{short=ML, long=machine-learning}
\DeclareAcronym{SegSNR}{short=SegSNR, long=segmental \ac{SNR}}
\DeclareAcronym{SegSSNR}{short=SegSSNR, long=segmental speech \ac{SNR}}
\DeclareAcronym{SegNR}{short=SegNR, long=segmental noise reduction}
\DeclareAcronym{MUSHRA}{short=MUSHRA, long=multi-stimulus test with hidden reference and anchor}
\DeclareAcronym{ANOVA}{short=ANOVA, long=analysis of variance}
\DeclareAcronym{MLSE}{short=MLSE, long=machine-learning spectral envelope}

\newcommand{\sExpect}{\mathbb{E}}

\newcommand{\sFreqIdx}{k}
\newcommand{\sFrameIdx}{\ell}

\newcommand{\sIdxPair}{{\sFreqIdx, \sFrameIdx}}

\newcommand{\sNoise}{n}
\newcommand{\sSpeech}{s}
\newcommand{\sNoisy}{y}

\newcommand{\sFreq}[1]{\MakeUppercase{#1}}

\newcommand{\sLog}[1]{\mathring{#1}}

\newcommand{\sPhase}{\Phi}
\newcommand{\sNoiseFreq}{\sFreq{\sNoise}}
\newcommand{\sNoiseFreqI}{\sNoiseFreq_{\sIdxPair}}

\newcommand{\sSpeechFreq}{\sFreq{\sSpeech}}
\newcommand{\sSpeechFreqIEst}{\sEst{\sSpeechFreq}_{\sIdxPair}}
\newcommand{\sSpeechFreqI}{\sSpeechFreq_{\sIdxPair}}

\newcommand{\sSpeechMagFreq}{A}

\newcommand{\sSpeechMagFreqIEst}{\sEst{\sSpeechMagFreq}_{\sIdxPair}}
\newcommand{\sNoisyFreq}{\sFreq{\sNoisy}}
\newcommand{\sNoisyFreqI}{\sNoisyFreq_{\sIdxPair}}
\newcommand{\sNoisyPhsFreq}{\sPhase^\sNoisy_{\sIdxPair}}

\newcommand{\sSpeechLog}{\sLog{\sSpeech}}

\newcommand{\sSpeechLogI}{\sSpeechLog_{\sIdxPair}}

\newcommand{\sVec}[1]{\mathbf{#1}}
\newcommand{\sEst}[1]{\hat{#1}}

\newcommand{\sCovFreq}{\Lambda}
\newcommand{\sVarNoisyFreq}{\sCovFreq^{\sNoisy}_\sIdxPair}
\newcommand{\sVarNoiseFreq}{\sCovFreq^{\sNoise}_\sIdxPair}
\newcommand{\sVarSpeechFreq}{\sCovFreq^{\sSpeech}_\sIdxPair}
\newcommand{\sVarSpeechFreqSGM}{\sCovFreq^{\sSpeech|\sState}_\sFreqIdx}

\newcommand{\sVarNoiseFreqEst}{\sEst{\sCovFreq}^{\sNoise}_\sIdxPair}
\newcommand{\sVarSpeechFreqEst}{\sEst{\sCovFreq}^{\sSpeech}_\sIdxPair}

\newcommand{\sCompression}{\beta} \newcommand{\sShape}{\mu}

\newcommand{\sPDF}{f}

\newcommand{\sGainI}{G_{\sIdxPair}}

\newcommand{\sIteratorA}{i}

\newcommand{\sState}{q}

\newcommand{\sNumStates}{Q}

\newcommand{\sPriorSNR}{\xi}
\newcommand{\sPriorSNRI}{\sPriorSNR_{\sIdxPair}}
\newcommand{\sPriorSNRIEst}{\sEst{\sPriorSNR}_{\sIdxPair}}

\newcommand{\sPostSNRI}{\gamma_{\sIdxPair}}
\newcommand{\sEphraimSubst}{\zeta}
\newcommand{\sEphraimSubstI}{\sEphraimSubst_{\sIdxPair}}

\newcommand{\sHypergeom}{\mathcal{M}}
\newcommand{\sGamma}{\Gamma}

\newcommand{\sFeature}{v}

\newcommand{\sFeatureVec}{\sVec{\sFeature}}

\newcommand{\sHiddenOutput}{h}
\newcommand{\sNumHiddenOutputs}{H}

\newcommand{\sDNNOutputIter}{j}
\newcommand{\sFeatureDim}{V}

\newcommand{\sPerMat}{\sVec{Y}}

\newcommand{\sBaseEl}{b}

\newcommand{\sBaseMat}{\sVec{\MakeUppercase{\sBaseEl}}}

\newcommand{\sBaseMatS}{\sBaseMat^{(\sSpeech)}}
\newcommand{\sBaseMatN}{\sBaseMat^{(\sNoise)}}

\newcommand{\sActiEl}{h}

\newcommand{\sActiMat}{\sVec{\MakeUppercase{\sActiEl}}}

\newcommand{\sActiMatS}{\sActiMat^{(\sSpeech)}}
\newcommand{\sActiMatN}{\sActiMat^{(\sNoise)}}

\newcommand{\sNumBases}{I}
\newcommand{\sNumBasesS}{\sNumBases^{(\sSpeech)}}
\newcommand{\sNumBasesN}{\sNumBases^{(\sNoise)}}

\newcommand{\sFrameSetState}{\mathbb{L}^{(\sState)}}

\newcommand{\sSparsity}{\nu}
\newcommand{\sSigLevel}{\alpha}
\newcommand{\sCost}{C}

\AddToShipoutPicture{%
    \put(50,20){%
        \parbox{1\textwidth}{%
            \begin{center}
                \footnotesize%
                This is the accepted version of the publication.
                The original publication is available at \url{https://doi.org/10.1109/TASLP.2017.2778151}.\\
                2329-9290 (c) 2017 IEEE. Personal use is permitted, but republication/redistribution requires IEEE permission.
                See \url{http://www.ieee.org/publications_standards/publications/rights/index.html} for more information.
            \end{center}
        }
    }%
}

\begin{document}

\title{On the Importance of Super-Gaussian Speech Priors for Machine-Learning Based Speech Enhancement}

\author{Robert~Rehr and
        Timo~Gerkmann% <-this % stops a space
        \thanks{R. Rehr and T. Gerkmann are with the Computer Science Department, Signal Processing (SP), Universität Hamburg, Germany (e-mail: robert.rehr@uni-hamburg.de; timo.gerkmann@uni-hamburg.de)}% <-this % stops a space
}

\markboth{}%
{}

\maketitle

\begin{abstract}
    For enhancing noisy signals, machine-learning based single-channel speech enhancement schemes exploit prior knowledge about typical speech spectral structures.
To ensure a good generalization and to meet requirements in terms of computational complexity and memory consumption, certain methods restrict themselves to learning speech spectral envelopes.
We refer to these approaches as \ac{MLSE}-based approaches.

In this paper we show by means of theoretical and experimental analyses that for \ac{MLSE}-based approaches, super-Gaussian priors allow for a reduction of noise between speech spectral harmonics which is not achievable using Gaussian estimators such as the Wiener filter. 
For the evaluation, we use a \ac{DNN}-based phoneme classifier and a low-rank \ac{NMF} framework as examples of \ac{MLSE}-based approaches.
A listening experiment and instrumental measures confirm that while super-Gaussian priors yield only moderate improvements for classic enhancement schemes, for \ac{MLSE}-based approaches super-Gaussian priors clearly make an important difference and significantly outperform Gaussian priors.


\acresetall%

\end{abstract}

\IEEEpeerreviewmaketitle

\section{Introduction}

\IEEEPARstart{I}{n} the presence of background noise, speech may be corrupted such that the perceived quality and possibly also the intelligibility are deteriorated.
Similarly, also human-machine interaction by means of automatic speech recognition systems may suffer from additional background noises.
Hence, the enhancement of corrupted speech signals is an important task for many applications, e.g., in telecommunications, for speech recognition, and for hearing aids.
In this paper, we consider single-channel methods that either assume that the noisy speech signal has been captured by a single microphone or process the output of a beamformer.

Single-channel speech enhancement has been a topic of research for decades and has given rise to many different approaches, e.g.,~\cite{breithaupt_parameterized_2008, krawczyk-becker_mmse-based_2016, ephraim_speech_1984, ephraim_speech_1985, breithaupt_novel_2008, gerkmann_noise_2011}.
Many approaches are formulated in the \ac{STFT} domain where a multiplicative gain function is applied to the complex spectra to suppress the bands which mainly contain noise.
A common approach is to estimate the clean speech coefficients blindly from the noisy observation.
For this, many different approaches have been proposed in the literature, e.g.,~\cite{ephraim_speech_1984, ephraim_speech_1985, you_s-order_2005, andrianakis_mmse_2006, erkelens_minimum_2007, breithaupt_parameterized_2008, hendriks_log-spectral_2009, hendriks_dft-domain_2013}.
These methods often require an estimate of the speech \ac{PSD} and the noise \ac{PSD} which are also estimated blindly from the noisy observation, e.g., using~\cite{martin_noise_2001, ephraim_speech_1984, breithaupt_novel_2008, gerkmann_noise_2011}.
These methods generally track the speech and noise \acp{PSD} over time, i.e., a time-varying estimate is returned.
Special attention has been turned to super-Gaussian clean speech estimators~\cite{you_s-order_2005, andrianakis_mmse_2006, erkelens_minimum_2007, breithaupt_parameterized_2008, hendriks_log-spectral_2009, krawczyk-becker_mmse-based_2016} as studies indicate that the complex Fourier coefficients are rather super-Gaussian than Gaussian distributed~\cite{martin_speech_2005, lotter_speech_2005}.

Another approach to estimate the clean speech \ac{PSD} and possibly also the noise \ac{PSD} is to employ \ac{ML} based methods, where the structure of speech and noise is learned before the processing takes place.
In this paper, a specific type of \ac{ML}-based algorithm is considered, where the learned speech models only represent the spectral envelope, e.g.,~\cite{ephraim_bayesian_1992, burshtein_speech_2002, srinivasan_codebook_2006, zhao_hmm-based_2007, yoshioka_speech_2011, chazan_hybrid_2016, he_multiplicative_2017}.
This means that harmonic structures caused by the vibrating vocal cords are not included.
This increases the generalizability and also reduces the computational complexity and the amount of data required for training.
This type of enhancement is referred to as \ac{MLSE}-based in this paper.
Contrarily, to distinguish this type of enhancement schemes from the classic estimation schemes considered above, we refer to the latter as non-\ac{MLSE}-based enhancement schemes.
While \ac{MLSE} approaches exploit prior knowledge about typical speech spectral structures, the envelope representation also limits the quality of the enhanced signal.
Due to the coarse representation of speech, residual noise may remain especially between spectral harmonics.
To reduce the undesired residual noise between harmonics, different solutions have been proposed.
In~\cite{yoshioka_speech_2011}, a harmonic model has been used to attenuate the remaining noise component between harmonics.
Contrarily, an estimate of the speech presence probability is employed in~\cite{chazan_hybrid_2016, he_multiplicative_2017} to attain a suppression of the residual noise.

In this paper, we show that if super-Gaussian clean speech estimators are used, postprocessing as in~\cite{yoshioka_speech_2011, chazan_hybrid_2016, he_multiplicative_2017} is not necessary.
For this, we consider the parameterized clean speech estimator proposed in~\cite{breithaupt_parameterized_2008}.
An analysis of this estimator shows that, under a super-Gaussian speech model, the background noise can be reduced even if the speech \ac{PSD} is overestimated, e.g., between spectral harmonics when modeling only the envelope.
Furthermore, the estimator in~\cite{breithaupt_parameterized_2008} is employed in two \ac{MLSE}-based enhancement schemes.
Both methods serve as examples and can be considered as variants of previously proposed methods in the literature.
The first one is a \ac{DNN}-based scheme similar to~\cite{chazan_hybrid_2016} which is chosen due to its similarities to other \ac{MLSE}-based enhancement methods, e.g.,~\cite{ephraim_bayesian_1992, burshtein_speech_2002, srinivasan_codebook_2006, zhao_hmm-based_2007, yoshioka_speech_2011, he_multiplicative_2017}.
To demonstrate the effectiveness of super-Gaussian estimators also for other \ac{MLSE}-based enhancement methods, the estimator in~\cite{breithaupt_parameterized_2008} is additionally embedded in a supervised, sparse \ac{NMF} enhancement scheme based on~\cite{fevotte_nonnegative_2008, le_roux_sparse_2015}.
Here, a low amount of basis vectors is employed such that mainly spectral envelopes are represented by the \ac{NMF} basis vectors.
We show that for the used non-\ac{MLSE}-based enhancement scheme, which is capable of estimating the spectral fine structure of speech, the super-Gaussian speech model yields only small improvements.
However, for the considered \ac{MLSE}-based enhancement schemes, where only a speech envelope model is employed, the super-Gaussian model has a very beneficial effect as it allows to remove disturbing residual noises.
Besides the \ac{MLSE} approaches addressed here, also \ac{MLSE} approaches with LogMax (also known as MixMax) mixing models benefit from this effect~\cite{rehr_mixmax_2017}.
Super-Gaussian speech models have also been previously employed in \ac{ML} based speech enhancement algorithms, e.g.,~\cite{hao_speech_2010, mohammadiha_spectral_2013, aroudi_hidden_2015}.
However, none of the papers provides an explicit analysis of the obtained improvements over Gaussian estimators in terms of the gain functions that result under super-Gaussian speech models.
Furthermore, the advantages of these estimators in combination with spectral speech envelope models have not been highlighted.

The paper is structured as follows.
First, we recapitulate the clean speech estimator proposed in~\cite{breithaupt_parameterized_2008} in Section~\ref{sec:SignalModel}.
After that, we describe the considered \ac{MLSE}-based enhancement schemes in Section~\ref{sec:PreTrained:DNN} and Section~\ref{sec:PreTrained:NMF}.
In Section~\ref{sec:Analysis} and Section~\ref{sec:Evaluation}, an analysis of the super-Gaussian estimator~\cite{breithaupt_parameterized_2008} and, respectively, a comparison of clean speech estimators employed in different enhancement schemes is presented.
In Section~\ref{sec:SubjectiveEvaluation}, the results of the subjective evaluation test are reported.

\section{Signal Model and Speech Estimators}
\label{sec:SignalModel}

In this section, we revisit the clean speech estimator~\cite{breithaupt_parameterized_2008}.
This estimator is parameterized such that various known estimators, e.g.,~\cite{ephraim_speech_1984, ephraim_speech_1985, you_s-order_2005, andrianakis_mmse_2006, hendriks_log-spectral_2009}, result as special cases.
In particular, it allows to incorporate super-Gaussian speech models and the estimation of compressed amplitudes.
As in~\cite{breithaupt_analysis_2011}, we use the name \ac{MOSIE} for the estimator in~\cite{breithaupt_parameterized_2008}.

In this paper, we employ input signals with a sampling rate of 16~kHz.
As \ac{MOSIE} operates in the \ac{STFT}-domain, the sampled noisy input signal is split into overlapping segments and each segment is transformed to the Fourier domain after an analysis window has been applied.
The segment length of the \ac{STFT} is set to 32~ms and a segment overlap of 50~\% is employed.
This yields the noisy spectra $\sNoisyFreqI$, where $\sFreqIdx$ denotes the frequency index and $\sFrameIdx$ the segment index.
The physically plausible additive corruption model is used where the noisy coefficients $\sNoisyFreqI$ are described as
\begin{equation}
    \label{eq:AdditiveMixing}
    \sNoisyFreqI = \sSpeechFreqI + \sNoiseFreqI.
\end{equation}
Here, $\sSpeechFreqI$ and $\sNoiseFreqI$ represent the clean speech and noise spectral coefficients, respectively.
The estimate of the clean speech spectral coefficients~$\sSpeechFreqIEst$ is obtained from the noisy observation~$\sNoisyFreqI$ using~\cite{breithaupt_parameterized_2008}.
Afterwards, the estimated clean speech spectra~$\sSpeechFreqIEst$ are transformed back to the time-domain and a synthesis window is applied to the obtained time-domain segments.
For the analysis and the synthesis a square-root Hann window is used.
Finally, an overlap-add method is used to reconstruct the complete time-domain signal.
The \ac{STFT} framework is shared among all enhancement schemes including the \ac{DNN}-based scheme in Section~\ref{sec:PreTrained:DNN} and the \ac{NMF}-based scheme in Section~\ref{sec:PreTrained:NMF}.

\ac{MOSIE}~\cite{breithaupt_parameterized_2008} is a statistically optimal estimator in the sense of the \ac{MSE}.
Such estimators consider the quantities in~\eqref{eq:AdditiveMixing} as random variables, where the involved \acp{PDF} are assumed to be known.
In~\cite{breithaupt_parameterized_2008}, the estimate $\sSpeechFreqIEst$ that minimizes the \ac{MSE} given by $\sExpect\{(|\sSpeechFreqI|^\sCompression - |\sSpeechFreqIEst|^\sCompression)^2\}$ has been derived.
Here, $\sExpect\{\cdot\}$ denotes the expectation operator and $|\cdot|^\sCompression$ allows to incorporate perceptually motivated compression functions.
Here, $\sCompression$ denotes the compression factor.
In general, the \ac{MSE} optimal estimator of $\sSpeechFreqI$ depends on the \acp{PDF} of the speech spectral coefficients $\sSpeechFreqI$ and the noise spectral coefficients $\sNoiseFreqI$.

In~\cite{breithaupt_parameterized_2008}, the complex noise coefficients~$\sNoiseFreqI$ are assumed to follow a circular-symmetric complex Gaussian distribution.
This assumption is often motivated by the Fourier sum and the central limit theorem~\cite{lotter_speech_2005}.
However, due to the strong correlation of speech in the time-domain, a Gaussian distribution does not appropriately describe the speech spectral coefficients $\sSpeechFreqI$~\cite{martin_speech_2005, lotter_speech_2005, erkelens_minimum_2007}.
Accordingly, a parametrizable circular-symmetric possibly heavy-tailed super-Gaussian distribution is employed to describe $\sSpeechFreqI$ in~\cite{breithaupt_parameterized_2008}.
Given the mixing model in~\eqref{eq:AdditiveMixing} and the statistical assumptions about the noise and speech coefficients, the estimate of the amplitude $\sSpeechMagFreqIEst$ is given by~\cite{breithaupt_parameterized_2008}
\begin{equation}
    \label{eq:MOSIE}
    \sSpeechMagFreqIEst = \sqrt{\frac{\sVarNoiseFreq \sPriorSNRI}{\sPriorSNRI + \sShape}}
    {\left[\frac{\sGamma(\sShape + \sCompression/2)}{\sGamma(\sShape)} \frac{\sHypergeom(\sShape + \sCompression/2, 1; \sEphraimSubstI)}{\sHypergeom(\sShape,1; \sEphraimSubstI)} \right]}^{\frac{1}{\sCompression}}.
\end{equation}
Here, $\sPriorSNRI = \sVarSpeechFreq / \sVarNoiseFreq$ denotes the \emph{a priori} \ac{SNR}.
The quantities $\sVarSpeechFreq = \sExpect\{|\sSpeechFreqI|^2\}$ and $\sVarNoiseFreq = \sExpect\{|\sNoiseFreqI|^2\}$ are the speech \ac{PSD} and the noise \ac{PSD}, respectively.
Further, $\sEphraimSubstI$ is given by $\sPostSNRI \sPriorSNRI / (\sShape + \sPriorSNRI)$ where $\sPostSNRI = |\sNoisyFreqI|^2 /  \sVarNoiseFreq$ is the \emph{a posteriori} \ac{SNR}.
The symbol $\sHypergeom(\cdot, \cdot; \cdot)$ represents the confluent hypergeometric function.
The parameter $\sShape > 0$ determines the shape of speech prior \ac{PDF} where $\sShape < 1$ corresponds to a super-Gaussian distribution while $\sShape = 1$ corresponds to a Gaussian distribution.
To obtain an estimate of the complex speech coefficients $\sSpeechFreqIEst$, the estimated amplitude in~\eqref{eq:MOSIE} is combined with the noisy phase $\sNoisyPhsFreq$ as $\sSpeechFreqIEst = \sSpeechMagFreqIEst \exp(j \sNoisyPhsFreq)$, where $j = \sqrt{-1}$.

It is interesting to note that \ac{MOSIE}~\cite{breithaupt_parameterized_2008}, generalizes existing clean speech estimators.
For example, if $\sCompression = 1$ and $\sShape = 1$, \ac{MOSIE}~\cite{breithaupt_parameterized_2008} is equivalent to Ephraim and Malah's \ac{STSA}~\cite{ephraim_speech_1984} and, for very small values of $\sCompression$ and $\sShape = 1$, the \ac{LSA}~\cite{ephraim_speech_1985} is approximated.
Super-Gaussian estimators are obtained for $\sShape < 1$.
Table~\ref{tab:RelatedEstimators} gives an overview over the related estimators.

\begin{table}[bt]
    \centering
    \caption{\label{tab:RelatedEstimators}List of clean speech estimators that \ac{MOSIE}~\cite{breithaupt_parameterized_2008} generalizes.}
    \begin{tabular}{ccc}
        \toprule
        $\sShape$      & $\sCompression$               & Related estimator\\
        \midrule
         1             & 1                             & Gaussian \acs{STSA}~\cite{ephraim_speech_1984}\\
         1             & $\sCompression \rightarrow 0$ & Gaussian \acs{LSA}~\cite{ephraim_speech_1985}\\
         $\sShape < 1$ & 1                             & super-Gaussian \ac{STSA}~\cite{erkelens_minimum_2007, andrianakis_mmse_2006}\\
         $\sShape < 1$ & $\sCompression \rightarrow 0$ & super-Gaussian \ac{LSA}~\cite{hendriks_log-spectral_2009}\\
        \bottomrule
    \end{tabular}
\end{table}

To evaluate the expression in~\eqref{eq:MOSIE}, estimates of the speech \ac{PSD} $\sVarSpeechFreq$ and the noise \ac{PSD} $\sVarNoiseFreq$ are required.
These can be obtained from non-\ac{MLSE}-based speech \ac{PSD} and noise \ac{PSD} estimators.
In this paper, the noise \ac{PSD}~$\sVarNoiseFreq$ is estimated using~\cite{gerkmann_noise_2011}.
The speech \ac{PSD} of the non-\ac{MLSE}-based enhancement scheme is estimated using temporal cepstrum smoothing as proposed in~\cite{breithaupt_novel_2008}.
The enhancement scheme that results from using these speech and noise \ac{PSD} estimator in \ac{MOSIE} is referred to as non-\ac{MLSE}-based enhancement scheme throughout this paper.
However, also \ac{ML} based estimators of the clean speech and the noise \ac{PSD} can be employed which are considered next.

\section{DNN-Based Speech Enhancement Scheme}
\label{sec:PreTrained:DNN}

As the first example of an \ac{MLSE} enhancement scheme, a method using a \ac{DNN}-based phoneme recognizer similar to~\cite{chazan_hybrid_2016} is considered.
Similarly, \ac{MLSE} models have also been used for enhancement schemes in~\cite{ephraim_bayesian_1992, burshtein_speech_2002, srinivasan_codebook_2006, zhao_hmm-based_2007, yoshioka_speech_2011, he_multiplicative_2017}.
In~\cite{chazan_hybrid_2016}, a two step procedure is used for speech enhancement.
First, the spoken phoneme is identified from the noisy observation.
After that, a learned speech \ac{PSD} corresponding to the recognized phoneme is used in a clean speech estimator, e.g., \ac{MOSIE}~\cite{breithaupt_parameterized_2008}, to enhance the noisy observation.
As speech is modeled on a phoneme level, the speech spectral fine structures, e.g., the spectral harmonics, are not resolved.

\begin{figure}[tb]
    \centering
    \newcommand{\relu}{\tiny ReLU}
\newcommand{\softmax}{\tiny soft\-max} 

\scalebox{0.9}{%
\begin{tikzpicture}[
        every node/.style={node distance=10pt and 35pt, font=\footnotesize},
        input/.style={%
            draw,
            fill=lightgray,
        },
        neuron/.style={circle, 
            draw, 
            font=\scriptsize, 
            inner sep=1.5pt, 
            text width=17pt, 
            align=center, 
            minimum width=22pt},
    ]

    \node[input, label=above:$\sFeature_{1, \sFrameIdx}$]                    (input1) {};

    \node[neuron, label=above:$\sHiddenOutput_{1, 1}$, right=of input1]        (hidden11) {\relu};
    \node[neuron, label=below:$\sHiddenOutput_{1, 2}$, below=of hidden11] (hidden12) {\relu};
    \node[below=of hidden12] (hidden13) {$\vdots$};
    \node[neuron, label=below:$\sHiddenOutput_{1, \sNumHiddenOutputs_1}$, below=of hidden13] (hidden1N) {\relu};

    \node[input, label=left:$\sFeature_{2, \sFrameIdx}$, left=of hidden12]   (input2) {};
    \node[left=40pt of hidden13] (input3) {$\vdots$};
    \node[input, label=below:$\sFeature_{\sFeatureDim, \sFrameIdx}$, left=of hidden1N] (inputN) {};

    \node[neuron, label=above:$\sHiddenOutput_{2, 1}$, right=of hidden11]        (hidden21) {\relu};
    \node[neuron, label=below:$\sHiddenOutput_{2, 2}$, below=of hidden21] (hidden22) {\relu};
    \node[below=of hidden22] (hidden23) {$\vdots$};
    \node[neuron, label=below:$\sHiddenOutput_{2, \sNumHiddenOutputs_2}$, below=of hidden23] (hidden2N) {\relu};

    \node[neuron, label=above:{$\sPDF(\sState = 1 | \sFeatureVec_\sFrameIdx)$}, right=of hidden21]        (hidden31) {\softmax};
    \node[neuron, label=right:{$\sPDF(\sState = 2 | \sFeatureVec_\sFrameIdx)$}, below=of hidden31] (hidden32) {\softmax};
    \node[below=of hidden32] (hidden33) {$\vdots$};
    \node[neuron, label=below:{$\sPDF(\sState = \sNumStates | \sFeatureVec_\sFrameIdx)$}, below=of hidden33] (hidden3N) {\softmax};

    \node[above=of input1, yshift=2.5ex] {Input layer};
    \node[above=of hidden11, yshift=0.75ex] {Hidden layer 1};
    \node[above=of hidden21, yshift=0.75ex] {Hidden layer 2};
    \node[above=of hidden31, yshift=0.75ex] {Output layer};

    \foreach \i in {1,2,N} {%
        \foreach \j in {1, 2, N} {%
            \draw[-latex] (input\i) -- (hidden1\j);
        }
    }

    \foreach \p/\n in {1/2, 2/3} {%
        \foreach \i in {1,2,N} {%
            \foreach \j in {1, 2, N} {%
                \draw[-latex] (hidden\p\i) -- (hidden\n\j);
            }
        }
    }
\end{tikzpicture}}%
\\
    \caption{\label{fig:DNNArchitecture}Architecture of the employed \acs{DNN}.}
\end{figure}
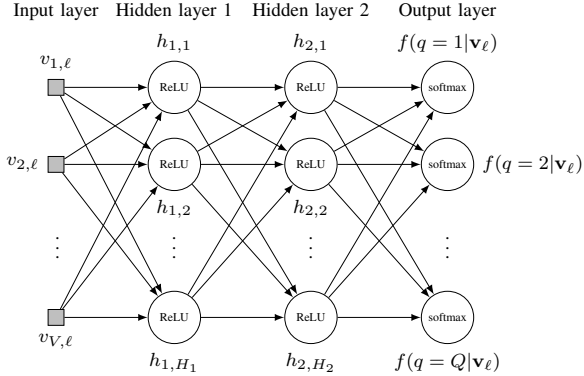

For phoneme recognition, a \ac{DNN} is used with the architecture shown in Fig.~\ref{fig:DNNArchitecture}.
The \ac{DNN}'s input is given by 13~\acp{MFCC} including the $\Delta$ and $\Delta\Delta$ accelerations which are extracted for each frame~$\sFrameIdx$.
To these features, a context is added by including the features of the three previous and three future segments which results in the feature vector $\sFeatureVec_\sFrameIdx = {[\sFeature_{1, \sFrameIdx}, \ldots, \sFeature_{\sFeatureDim, \sFrameIdx}]}^T$ with dimensionality $\sFeatureDim = 273$.
Here, $\sFeature_{\sIteratorA, \sFrameIdx}$ denote the elements of the feature vector $\sFeatureVec_\sFrameIdx$.
Further, $\cdot^T$ denotes the vector and matrix transpose.
For the employed segment length and segment shift, the context is approximately 100~ms.
To improve the robustness of the recognition in noisy environments, the feature vectors are normalized using \ac{CMVN}~\cite{viikki_cepstral_1998} before they are employed for training or testing~\cite{chazan_hybrid_2016}.
The \ac{CMVN} is applied per utterance.

The features are passed through two hidden layers to finally obtain a score $\sPDF(\sState | \sFeatureVec_\sFrameIdx)$ for each phoneme $\sState \in \{1,\ldots,\sNumStates\}$.
We base the number of phonemes on the annotation given in the TIMIT database~\cite{garofolo_timit_1993} which distinguishes between $\sNumStates = 61$ classes including pauses and non-speech events. 
The hidden layers of the \ac{DNN} consist of $\sNumHiddenOutputs_1$ and $\sNumHiddenOutputs_2$ outputs, where $\sNumHiddenOutputs_1 = \sNumHiddenOutputs_2 = 512$ is used.
Similar to~\cite{toth_phone_2013, dahl_improving_2013, chazan_hybrid_2016}, \acp{ReLU} are employed as transfer functions of these two layers.
For the output layer, a softmax transfer function is used which is interpreted as the posterior probability $\sPDF(\sState|\sFeatureVec_\sFrameIdx)$ that phoneme~$\sState$ was spoken given the features $\sFeatureVec_\sFrameIdx$.

\begin{algorithm}[tb]
    \begin{algorithmic}[1]
        \REQUIRE%
        Trained \ac{DNN} and offline computed $\sVarSpeechFreqSGM$.
        \REQUIRE%
        Noisy observations $\sNoisyFreqI$ of a complete utterance.
        \STATE%
        Extract \acp{MFCC} $\sFeatureVec_\sFrameIdx$ from $\sNoisyFreqI$ for complete utterance and add context.
        \STATE%
        Apply \ac{CMVN} over complete utterance to give $\sFeatureVec_\sFrameIdx$.\\
        \FORALL{segments $\sFrameIdx$}
        \STATE%
        Estimate noise \ac{PSD} $\sVarNoiseFreqEst$ using~\cite{gerkmann_noise_2011}.
        \STATE%
        Obtain $\sPDF(\sState|\sFeatureVec_\sFrameIdx)$ from the \ac{DNN}.
        \FORALL{phonemes $\sState$}
        \STATE%
        Obtain clean speech estimate $\sSpeechFreqIEst^{(\sState)}$ for phoneme $\sState$.\\
        For this, $\sVarSpeechFreqSGM$ and $\sVarNoiseFreqEst$ are employed in~\eqref{eq:MOSIE}.
        \ENDFOR%
        \STATE%
        Obtain final clean speech estimate $\sSpeechFreqIEst$ using~\eqref{eq:DNN:Combination}.
        \ENDFOR%
    \end{algorithmic}
    \caption{\label{alg:DNN}\acs{DNN}-based enhancement scheme.}
\end{algorithm}

For the enhancement, \ac{MLSE}-based clean speech \acp{PSD} $\sVarSpeechFreqSGM$ are employed where each $\sVarSpeechFreqSGM$ represents the speech \ac{PSD} of a specific phoneme~$\sState$.
During processing, each $\sVarSpeechFreqSGM$ is used in~\eqref{eq:MOSIE} via $\sPriorSNRI = \sVarSpeechFreqSGM / \sVarNoiseFreq$, which yields the phoneme specific clean speech estimates $\sSpeechFreqIEst^{(\sState)}$.
For this, the noise \ac{PSD}~$\sVarNoiseFreq$ is estimated using~\cite{gerkmann_noise_2011}.
Similar to~\cite{chazan_hybrid_2016}, the estimates $\sSpeechFreqIEst^{(\sState)}$ are averaged based on the recognition scores $\sPDF(\sState|\sFeatureVec_\sFrameIdx)$ to give a final estimate $\sSpeechFreqIEst$.
More specifically, the clean speech coefficients are obtained by
\begin{equation}
    \label{eq:DNN:Combination}
    \sSpeechFreqIEst = \sum_{\sDNNOutputIter = 1}^\sNumStates \sPDF(\sState = \sDNNOutputIter|\sFeatureVec_\sFrameIdx) \sSpeechFreqIEst^{(\sState)}.
\end{equation}
The steps required to enhance the noisy observations $\sNoisyFreqI$ using the \ac{DNN}-based enhancement scheme are summarized in Algorithm~\ref{alg:DNN}.

For the training of the \ac{DNN}-based \ac{MLSE} system, we employ 1196~gender and phonetically balanced sentences from the TIMIT training set.
As in~\cite{chazan_hybrid_2016}, the \ac{DNN} is trained only using clean speech to ensure that the phoneme recognition does not depend on the background noise type.
The target vectors for the training are given by a one-hot encoding of the TIMIT phoneme annotation~\cite{garofolo_timit_1993}.
The error function is given by the cross-entropy which is minimized using scaled conjugate gradient back-propagation~\cite{moller_scaled_1993}.
Before back-propagation, the weights of the \ac{DNN}'s two hidden layers are initialized using the Glorot method~\cite{glorot_understanding_2010}.
The weights of the output layer are initialized using the Nguyen-Widrow method~\cite{nguyen_improving_1990}.

Similar to the non-\ac{MLSE}-based enhancement scheme, the noise \ac{PSD}~$\sVarNoiseFreq$ is estimated using~\cite{gerkmann_noise_2011}.
The speech \acp{PSD}~$\sVarSpeechFreqSGM$ that are linked to the phonemes~$\sState$ are obtained as
\begin{equation}
    \label{eq:Train:SpeechPSD}
    \sVarSpeechFreqSGM = \frac{1}{|\sFrameSetState|} \sum_{\sFrameIdx \in \sFrameSetState} |\sSpeechFreqI|^2,
\end{equation}
where $\sFrameSetState$ denotes the set that contains the segments that belong to the phoneme~$\sState$ in the training data.
As~\eqref{eq:Train:SpeechPSD} is scale-dependent, we normalize the time-domain clean speech input signal both in training and testing such that all sentences have the same peak value.
During training, the clean speech data is available, while during testing, oracle knowledge is provided.
This normalization is also employed for the other enhancement schemes, i.e., for the non-\ac{MLSE}-based and the \ac{NMF}-based enhancement scheme given in Section~\ref{sec:PreTrained:NMF}.
Here, however, the normalization has no influence as these approaches are scale-independent.

\section{NMF-Based Speech Enhancement Scheme}
\label{sec:PreTrained:NMF}

In this part, the \ac{MLSE}-based enhancement scheme that employs \ac{NMF} is described.
It serves as a second example for \ac{MLSE}-based enhancement schemes.
\ac{NMF} approximates a nonnegative matrix~$\sPerMat$ as $\sPerMat \approx \sBaseMat \sActiMat$, where $\sBaseMat$ and $\sActiMat$ are also nonnegative matrices.
The columns of $\sBaseMat$ are referred to as basis vectors and the columns of $\sActiMat$ as activation vectors.
\ac{NMF} has been used for source separation, e.g.,~\cite{schmidt_single-channel_2006, virtanen_monaural_2007, fevotte_nonnegative_2008}, and has also been applied to speech enhancement, e.g.,~\cite{mohammadiha_new_2011, mohammadiha_nonnegative_2013, mohammadiha_supervised_2013}.

Here, a simple, supervised, sparse \ac{NMF} approach is used which employs the \ac{IS} divergence as the cost function~\cite{fevotte_nonnegative_2008, le_roux_sparse_2015}.
As argued in~\cite{fevotte_nonnegative_2008}, if the noisy spectral coefficients $\sNoisyFreqI$ are independent and follow a circular-symmetric Gaussian distribution, minimizing the \ac{IS} divergence for approximating the noisy periodogram as $\left[ |\sNoisyFreqI|^2 \right] =\sPerMat \approx \sBaseMat \sActiMat$ allows the elements of the product $\sBaseMat \sActiMat$ to be interpreted as the noisy \ac{PSD}~$\sVarNoisyFreq$.
The \ac{IS} cost function including the sparsity constraint is given by~\cite{le_roux_sparse_2015}
\begin{equation}
    \label{eq:NMF:CostFunction}
    \sCost = \sSparsity |\sActiMat|_1 + \sum_{i, j} \frac{(\sPerMat)_{i,j}}{(\sBaseMat \sActiMat)_{i, j}} + \log\left(\frac{(\sPerMat)_{i,j}}{(\sBaseMat \sActiMat)_{i, j}}\right) - 1,
\end{equation}
where $(\cdot)_{i, j}$ denotes element of the respective matrix, $|\cdot|_1$ the $L_1$-norm, and $\sSparsity$ is the factor that controls the sparsity.
This cost function can be optimized using the multiplicative update rules in~\cite{le_roux_sparse_2015}.

For estimating the speech and the noise \ac{PSD}, it is assumed that the basis matrix $\sBaseMat$ is given by the concatenation of a speech basis matrix $\sBaseMatS$ and a noise basis matrix $\sBaseMatN$ as $\sBaseMat = [\sBaseMatS, \sBaseMatN]$.
The speech and noise basis matrices are learned prior to the processing and are held fixed during processing.
This means that only the activation matrices are updated.
For obtaining an estimate of $\sVarSpeechFreq$ and $\sVarNoiseFreq$, also the activation matrix $\sActiMat$ is split into a speech and noise dependent part as $\sActiMat = {[(\sActiMatS)^T, (\sActiMatN)^T]}^T$ such that $\sPerMat \approx \sBaseMat \sActiMat = [\sBaseMatS, \sBaseMatN] {[(\sActiMatS)^T, (\sActiMatN)^T]}^T$.
With this, the speech and the noise \ac{PSD} can be obtained as
\begin{align}
    \label{eq:NMF:SpeechPSD}
    \sVarSpeechFreqEst &= \sum_{\sIteratorA = 1}^{\sNumBasesS} (\sBaseMatS)_{\sFreqIdx, \sIteratorA} (\sActiMatS)_{\sIteratorA, \sFrameIdx}\\
    \label{eq:NMF:NoisePSD}
    \sVarNoiseFreqEst &= \sum_{\sIteratorA = 1}^{\sNumBasesN} (\sBaseMatN)_{\sFreqIdx, \sIteratorA} (\sActiMatN)_{\sIteratorA, \sFrameIdx},
\end{align}
where $\sNumBasesS$ is the number of speech basis while $\sNumBasesN$ denotes the number of noise bases.
The steps for enhancing the noisy observations are summarized in Algorithm~\ref{alg:NMF}.

\begin{algorithm}[tb]
    \begin{algorithmic}[1]
        \REQUIRE%
        Speech and noise basis matrix $\sBaseMatS$, $\sBaseMatN$.
        \STATE%
        Set $\sBaseMat = [\sBaseMatS, \sBaseMatN]$.
        \FORALL{segments $\sFrameIdx$}
        \STATE%
        Create vector $\sVec{\sNoisy}_\sFrameIdx = |\sNoisyFreqI|^2$ and add context.
        \STATE%
        Initialize $\sActiMat$ with positive random numbers.
        \REPEAT%
        \STATE%
        Update $\sActiMat$ with the update rule in~\cite[(4)]{le_roux_sparse_2015}.
        \UNTIL{convergence or maximum iterations reached}
        \ENDFOR%
        \STATE%
        Obtain $\sVarSpeechFreqEst$ and $\sVarNoiseFreqEst$ using~\eqref{eq:NMF:SpeechPSD} and~\eqref{eq:NMF:NoisePSD}.
        \STATE%
        Use estimated \acp{PSD} in~\eqref{eq:MOSIE} to obtain $\sSpeechFreqIEst$.
    \end{algorithmic}
    \caption{\label{alg:NMF}\ac{NMF}-based enhancement scheme.}
\end{algorithm}

For the \ac{NMF}-based enhancement scheme, the same speech audio material is employed for training as for the \ac{DNN}-based enhancement scheme.
Also here, a context of 7~segments is employed, i.e., three past and three future segments are appended to the noisy input vectors.
As a consequence, the number of rows of the basis matrices is increased and the speech \ac{PSD} and the noise \ac{PSD} are reconstructed with a context.
For the enhancement, however, only the elements corresponding to the current segment are employed.
We use 30~bases in the speech basis matrix $\sBaseMatS$ and the noise basis matrix $\sBaseMatN$ while the sparsity weight in~\eqref{eq:NMF:CostFunction} is set to $\sSparsity = 10$.

The noise basis matrices~$\sBaseMatN$ are trained for a set of specific background noise types.
The used types are babble noise, factory~1 noise, and pink noise taken from the NOISEX-92 database~\cite{steeneken_description_1988}.
Further, an amplitude modulated version of the pink noise similar to~\cite{gerkmann_noise_2011} and a traffic noise taken from~\cite{fxprosound_audio_design_traffic_2009} are included.
These noise types are also used later in the evaluation in Section~\ref{sec:Evaluation}.
To ensure that different audio material is used in the evaluation, only the first two minutes of the respective noise type are used for training.
This corresponds to a partitioning where 50~\% of the background noise material is used for training and 50~\% for testing.
For training and testing, a maximum of 200~iterations are performed for the multiplicative updates in~\cite{le_roux_sparse_2015}.
For testing, the noise matrix appropriate for the respective noise type is chosen in the evaluation, i.e., the background noise type is assumed to be known.
The employed non-\ac{MLSE}-based and the \ac{DNN}-based enhancement scheme do not require such prior knowledge.
However, as discussed in~\cite{mohammadiha_supervised_2013, mohammadiha_state-space_2015}, such a supervised approach may be appropriate for some applications, e.g., where the environment can be identified using an environment classifier.

\section{Importance of Super-Gaussianity for MLSE Based Speech Enhancement}
\label{sec:Analysis}

In this section, we analyze the effect of the super-Gaussian speech estimators on non-\ac{MLSE}-based and \ac{MLSE}-based speech enhancement schemes.
Before that, we analyze how the shape $\sShape$ and the compression $\sCompression$ influence the behavior of \ac{MOSIE}~\cite{breithaupt_parameterized_2008}.

\subsection{Analysis of the Gain Functions}

In this part, we analyze the behavior of the clean speech estimator \ac{MOSIE}~\cite{breithaupt_parameterized_2008}.
For this, the gain function is considered which is defined as
\begin{align}
    \label{eq:GainFunction}
    \sGainI &= \sSpeechFreqIEst / \sNoisyFreqI\\
    \label{eq:GainFunctionAbs}
            &= |\sSpeechFreqIEst| / |\sNoisyFreqI|.
\end{align}
The equality between~\eqref{eq:GainFunction} and~\eqref{eq:GainFunctionAbs} holds due to the fact that \ac{MOSIE}~\cite{breithaupt_parameterized_2008} combines an estimate of the clean speech magnitude~$\sSpeechMagFreqIEst$ with the noisy phase~$\sNoisyPhsFreq$.
Thus, the gain is a real-valued function that describes by how much a spectral coefficient is boosted or attenuated depending on the speech \ac{PSD}~$\sVarSpeechFreq$, the noise \ac{PSD}~$\sVarNoiseFreq$, and the noisy input $\sNoisyFreqI$.

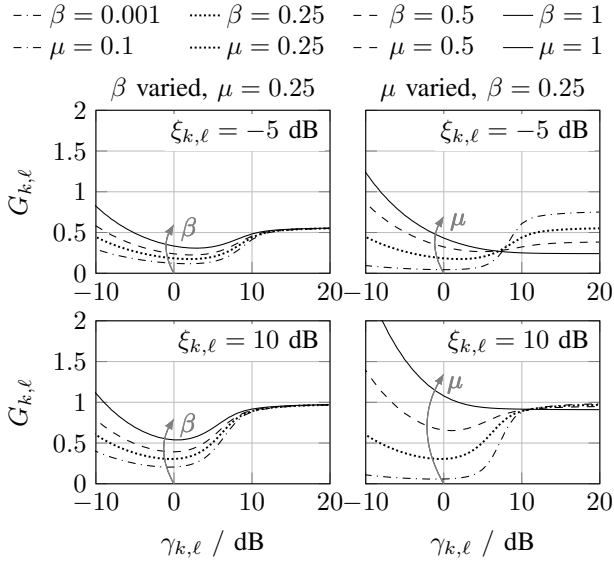
\begin{figure}[tb]
    \centering
    \begin{tikzpicture}
    \begin{customlegend}[
            legend entries={
                $\sCompression = 0.001$,
                $\sCompression = 0.25$,
                $\sCompression = 0.5$,
                $\sCompression = 1$~,
                $\sShape = 0.1$,
                $\sShape = 0.25$,
                $\sShape = 0.5$,
                $\sShape = 1$
            },
            legend columns=4,
            legend cell align={left},
            legend style={draw=none, inner sep=2pt, /tikz/every even column/.append style={column sep=0.3cm}},]
        ]
        \addlegendimage{comp0.0010}
        \addlegendimage{comp0.2500}
        \addlegendimage{comp0.5000}
        \addlegendimage{comp1.0000}
        \addlegendimage{shape0.10}
        \addlegendimage{shape0.25}
        \addlegendimage{shape0.50}
        \addlegendimage{shape1.00}
    \end{customlegend}
\end{tikzpicture}\\
    \newcounter{row}
\begin{tikzpicture}[
        snrlabel/.style={anchor=north east, fill=white, inner sep=2pt,},
    ]%
    \begin{groupplot}[
            group style={%
                group name=gain betaorder,
                group size=2 by 2,
                xlabels at=edge bottom,
                y descriptions at=edge left,
                vertical sep=4ex,
                horizontal sep=3ex,
            },
            width=0.53\linewidth,
            height=3.75cm,
            grid,
            xmin=-10,
            xmax=20,
            ymin=0,
            ymax=2,
            xlabel={$\sPostSNRI~/~\text{dB}$},
            ylabel={$\sGainI$},
        ]

        \edef\tmp{}
        \pgfplotsforeachungrouped \snr in {-5, 10} {%
            \eappto\tmp{\noexpand\nextgroupplot}
                \eappto\tmp{\noexpand\addlegendimage{comp1.0000}}
                \eappto\tmp{\noexpand\addlegendimage{comp0.5000}}
                \eappto\tmp{\noexpand\addlegendimage{comp0.2500}}
                \eappto\tmp{\noexpand\addlegendimage{comp0.0010}}
            \pgfplotsforeachungrouped \comp in {1.0000, 0.5000, 0.2500, 0.0010} {%
                \eappto\tmp{\noexpand\addplot[comp\comp] table[x=post_snr_db, y=gain] {Data/gain/gain_betaorder_\snr_0.25_\comp.tsv};}
                \ifnum\value{row} = 0%
                    \eappto\tmp{\noexpand\draw[latex-, gray] 
                    (axis cs:0, 0.6) to [bend right] node[pos=0.1, right] {$\sCompression$} (axis cs:0, 0);}
                \else
                    \eappto\tmp{\noexpand\draw[latex-, gray] 
                    (axis cs:0, 0.8) to [bend right] node[pos=0.1, right] {$\sCompression$} (axis cs:0, 0);}
                \fi
            }
            \eappto\tmp{\noexpand\node[snrlabel] at(rel axis cs:0.98, 0.98) {$\sPriorSNRI = \snr~\text{dB}$};}
            \eappto\tmp{\noexpand\nextgroupplot}
                \eappto\tmp{\noexpand\addlegendimage{shape1.00}}
                \eappto\tmp{\noexpand\addlegendimage{shape0.50}}
                \eappto\tmp{\noexpand\addlegendimage{shape0.25}}
            \pgfplotsforeachungrouped \shape in {1.00, 0.50, 0.25, 0.10} {%
                \eappto\tmp{\noexpand\addplot[shape\shape] table[x=post_snr_db, y=gain] {Data/gain/gain_betaorder_\snr_\shape_0.2500.tsv};}
                \ifnum\value{row} = 0%
                    \eappto\tmp{\noexpand\draw[latex-, gray] 
                    (axis cs:0, 0.7) to [bend right] node[pos=0.1, right] {$\sShape$} (axis cs:0, 0);}
                \else
                    \eappto\tmp{\noexpand\draw[latex-, gray] 
                    (axis cs:0, 1.35) to [bend right] node[pos=0.1, right] {$\sShape$} (axis cs:0, 0);}
                \fi
            }
            \eappto\tmp{\noexpand\node[snrlabel] at(rel axis cs:0.98, 0.98) {$\sPriorSNRI = \snr~\text{dB}$};}
            \stepcounter{row}
        }
        \tmp
    \end{groupplot}
    
    \node[anchor=south] at(gain betaorder c1r1.north) {$\sCompression$ varied, $\sShape = 0.25$};
    \node[anchor=south] at(gain betaorder c2r1.north) {$\sShape$ varied, $\sCompression = 0.25$};
\end{tikzpicture}%
\\
    \caption{\label{fig:MOSIEPost}Gain function~$\sGainI$ of \ac{MOSIE}~\cite{breithaupt_parameterized_2008} over the \emph{a posteriori} \acs{SNR}~$\sPostSNRI$ for different values of shape~$\sShape$ and compression~$\sCompression$.
        The upper row shows the results for an \emph{a priori}~\acs{SNR} of -5~dB and the lower for an \emph{a priori} \ac{SNR} of 10~dB.
        See Table~\ref{tab:RelatedEstimators} for related estimators for the values of $\sShape$ and $\sCompression$.
    }
\end{figure}

Fig.~\ref{fig:MOSIEPost} shows the gain~$\sGainI$ of \ac{MOSIE}~\cite{breithaupt_parameterized_2008} over the \emph{a posteriori} \ac{SNR} $\sPostSNRI$ for two \emph{a priori} \acp{SNR}: $\sPriorSNRI = -5~\text{dB}$ is shown in the upper row and $\sPriorSNRI = 10~\text{dB}$ in the lower row.
The compression parameter~$\sCompression$ is varied and the shape~$\sShape$ is kept fixed in the left panel and vice versa in the right panel.
It is well known that super-Gaussian estimators $(\sShape < 1)$ preserve speech better than Gaussian estimators $(\sShape = 1)$ for large \emph{a posteriori} \acp{SNR}~\cite{martin_speech_2005}.
However, in the context of \ac{MLSE}-based speech enhancement, it is of particular interest to observe in Fig.~\ref{fig:MOSIEPost} that with decreasing shape $\sShape$, a stronger attenuation is applied to the input coefficients for low \emph{a posteriori} \acp{SNR}~$\sPostSNRI$ even if the \emph{a priori} \ac{SNR} $\sPriorSNRI$ is large.
A similar effect is observed if a stronger compression, i.e., smaller values for~$\sCompression$, are employed.

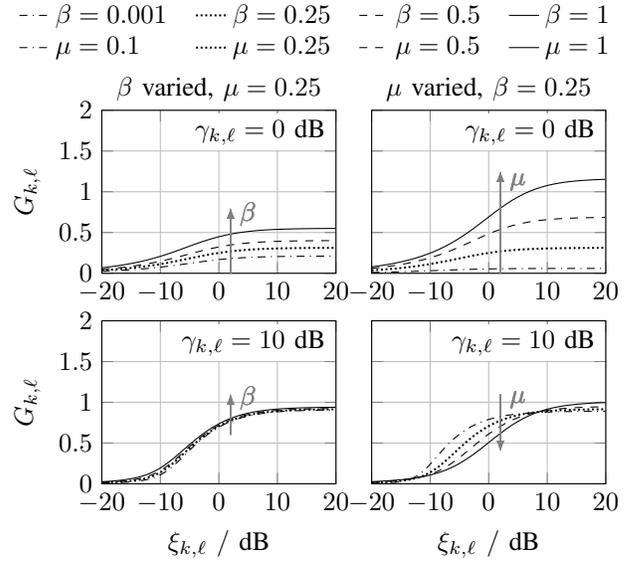
\begin{figure}[tb]
    \centering
    \begin{tikzpicture}
    \begin{customlegend}[
            legend entries={
                $\sCompression = 0.001$,
                $\sCompression = 0.25$,
                $\sCompression = 0.5$,
                $\sCompression = 1$~,
                $\sShape = 0.1$,
                $\sShape = 0.25$,
                $\sShape = 0.5$,
                $\sShape = 1$
            },
            legend columns=4,
            legend cell align={left},
            legend style={draw=none, inner sep=2pt, /tikz/every even column/.append style={column sep=0.3cm}},]
        ]
        \addlegendimage{comp0.0010}
        \addlegendimage{comp0.2500}
        \addlegendimage{comp0.5000}
        \addlegendimage{comp1.0000}
        \addlegendimage{shape0.10}
        \addlegendimage{shape0.25}
        \addlegendimage{shape0.50}
        \addlegendimage{shape1.00}
    \end{customlegend}
\end{tikzpicture}\\
    \setcounter{row}{0}
\begin{tikzpicture}[
        snrlabel/.style={anchor=north east, fill=white, inner sep=2pt,},
    ]%
    \begin{groupplot}[
            group style={%
                group name=gain betaorder,
                group size=2 by 2,
                xlabels at=edge bottom,
                y descriptions at=edge left,
                vertical sep=4ex,
                horizontal sep=3ex,
            },
            width=0.53\linewidth,
            height=3.75cm,
            grid,
            xmin=-20,
            xmax=20,
            ymin=0,
            ymax=2,
            xlabel={$\sPriorSNRI~/~\text{dB}$},
            ylabel={$\sGainI$},
        ]

        \edef\tmp{}
        \pgfplotsforeachungrouped \snr in {0, 10} {%
            \eappto\tmp{\noexpand\nextgroupplot}
                \eappto\tmp{\noexpand\addlegendimage{comp1.0000}}
                \eappto\tmp{\noexpand\addlegendimage{comp0.5000}}
                \eappto\tmp{\noexpand\addlegendimage{comp0.2500}}
                \eappto\tmp{\noexpand\addlegendimage{comp0.0010}}
            \pgfplotsforeachungrouped \comp in {1.0000, 0.5000, 0.2500, 0.0010} {%
                \eappto\tmp{\noexpand\addplot[comp\comp] table[x=prior_snr_db, y=gain] {Data/gain_apriori/gain_betaorder_\snr_0.25_\comp.tsv};}
                \ifnum\value{row} = 0%
                    \eappto\tmp{\noexpand\draw[latex-, gray] 
                    (axis cs:2, 0.8) to node[pos=0.1, right] {$\sCompression$} (axis cs:2, 0);}
                \else
                    \eappto\tmp{\noexpand\draw[latex-, gray] 
                    (axis cs:2, 1.1) to node[pos=0.1, right] {$\sCompression$} (axis cs:2, 0.6);}
                \fi
            }
            \eappto\tmp{\noexpand\node[snrlabel] at(rel axis cs:0.98, 0.98) {$\sPostSNRI = \snr~\text{dB}$};}
            \eappto\tmp{\noexpand\nextgroupplot}
                \eappto\tmp{\noexpand\addlegendimage{shape1.00}}
                \eappto\tmp{\noexpand\addlegendimage{shape0.50}}
                \eappto\tmp{\noexpand\addlegendimage{shape0.25}}
            \pgfplotsforeachungrouped \shape in {1.00, 0.50, 0.25, 0.10} {%
                \eappto\tmp{\noexpand\addplot[shape\shape] table[x=prior_snr_db, y=gain] {Data/gain_apriori/gain_betaorder_\snr_\shape_0.2500.tsv};}
                \ifnum\value{row} = 0%
                    \eappto\tmp{\noexpand\draw[latex-, gray] 
                    (axis cs:2, 1.25) to node[pos=0.1, right] {$\sShape$} (axis cs:2, 0);}
                \else
                    \eappto\tmp{\noexpand\draw[-latex, gray] 
                    (axis cs:2, 1.1) to  node[pos=0.1, right] {$\sShape$} (axis cs:2, 0.4);}
                \fi
            }
            \eappto\tmp{\noexpand\node[snrlabel] at(rel axis cs:0.98, 0.98) {$\sPostSNRI = \snr~\text{dB}$};}
            \stepcounter{row}
        }
        \tmp
    \end{groupplot}
    
    \node[anchor=south] at(gain betaorder c1r1.north) {$\sCompression$ varied, $\sShape = 0.25$};
    \node[anchor=south] at(gain betaorder c2r1.north) {$\sShape$ varied, $\sCompression = 0.25$};
\end{tikzpicture}%
    \caption{\label{fig:MOSIEPrior}Same as Fig.~\ref{fig:MOSIEPost} but over the \emph{a priori} \acs{SNR}~$\sPriorSNRI$ and for two fixed \emph{a posteriori}~\acsp{SNR} $\sPostSNRI = 0~\text{dB}$ and $\sPostSNRI = 10~\text{dB}$.
    }
\end{figure}

These observations are supported by Fig.~\ref{fig:MOSIEPrior} where the gain function~$\sGainI$ is shown over the \emph{a priori} \ac{SNR}~$\sPriorSNRI$.
Here, the two rows show the behavior for two \emph{a posteriori} \acp{SNR}~$\sPostSNRI = 0~\text{dB}$ and $\sPostSNRI = 10~\text{dB}$.
For the Gaussian case ($\sShape = 1$), Fig.~\ref{fig:MOSIEPrior} shows that the gain~$\sGainI$ mainly depends on the \emph{a priori} \ac{SNR}~$\sPriorSNRI$.
If the \emph{a posteriori} \ac{SNR}~$\sPostSNRI$ is close to 0~dB and low values for $\sCompression$ and $\sShape$ are employed, i.e., the super-Gaussian case is considered, the attenuation remains low over a wide range of \emph{a priori} \acp{SNR}~$\sPriorSNRI$.
Hence, for \ac{MLSE}-based speech enhancement schemes, the residual noise can be suppressed even for large overestimations of the \emph{a priori} \ac{SNR}~$\sPriorSNRI$.
This occurs, e.g., between speech spectral harmonics which are not resolved by spectral envelope models.

\subsection{Effects of Super-Gaussian Estimators on the Enhancement}
\label{sec:Analysis:Effects}

\begin{figure}[tb]
    \centering
    \begin{tikzpicture}
    \begin{axis}[
            width=0.75\linewidth,
            height=3.75cm,
            specgram_axis
        ]
        \getImage{Data/example/spectrogram_noisy.tsv}
    \end{axis}
\end{tikzpicture}\\
    \caption{\label{fig:Example:SpectrogramNoisy}Spectrogram of the example speech signal in stationary pink noise at at 5~dB \ac{SNR}.
        Here, $f$ denotes frequency and $t$ time.
    }
\end{figure}
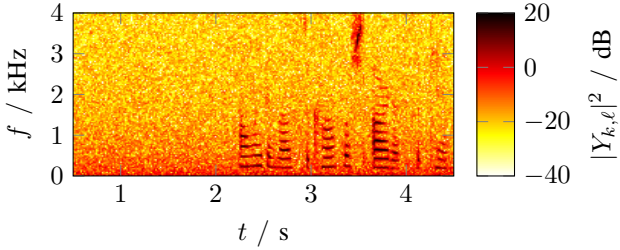

\begin{figure*}[tb]
    \centering
    \begin{tikzpicture}
    \begin{groupplot}[
            group style={%
                group size=3 by 1,
                xlabels at=edge bottom,
                ylabels at=edge left,
                vertical sep=4ex,
                horizontal sep=1.5em,
            },
            width=0.33\textwidth,
            height=3.75cm,
            priorsnr_axis,
        ]
        \edef\tmp{}

        \newcounter{gfxcount}
        \edef\colorbar{false}

        \pgfplotsforeachungrouped \method/\displayname in {generic/non-MLSE, nn/DNN, nmf/NMF} {%
            \ifnum\value{gfxcount} > 1
                \edef\colorbar{true}
            \fi
            \eappto\tmp{\noexpand\nextgroupplot[colorbar=\colorbar]}
            \eappto\tmp{\noexpand\getImage{Data/example/prior_snr_\method.tsv}}
            \eappto\tmp{\noexpand\node[anchor=north west, fill=white, rectangle, draw] at(rel axis cs: 0.01, 0.99) {\displayname};}
            \stepcounter{gfxcount}
        }
        \tmp
    \end{groupplot}
\end{tikzpicture}\\
    \caption{\label{fig:Example:PriorSNR}\emph{A priori} \ac{SNR}~$\sPriorSNRIEst$ estimated using different enhancement schemes for the excerpt shown in Fig.~\ref{fig:Example:SpectrogramNoisy}.
    Here, $f$ denotes frequency and $t$ time.
    }
\end{figure*}
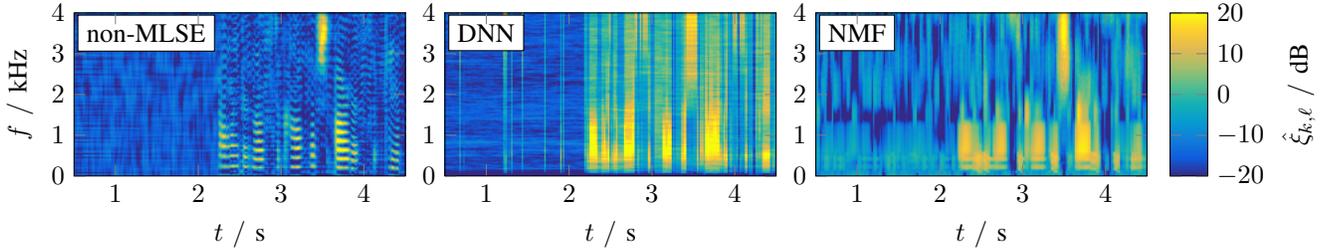

In this part, we analyze how the behavior of \ac{MOSIE}~\cite{breithaupt_parameterized_2008} influences the considered enhancement schemes.
For this, a speech signal taken from the TIMIT test set is corrupted by stationary pink noise at an \ac{SNR} of 5~dB.
The spectrogram of the used signal is shown in Fig.~\ref{fig:Example:SpectrogramNoisy}.
This signal is processed by the non-\ac{MLSE}-based enhancement scheme and the two \ac{MLSE}-based enhancement schemes.
In Fig.~\ref{fig:Example:PriorSNR}, we depict the resulting \emph{a priori} \acp{SNR}~$\sPriorSNRI$.
For the \ac{DNN}-based enhancement scheme, the \emph{a priori} \ac{SNR} of the phoneme that is most likely to be present is shown for each segment.
Note that this selection is only performed for the visualization in Fig.~\ref{fig:Example:PriorSNR}.
Otherwise, $\sSpeechFreqIEst$ is estimated as in~\eqref{eq:DNN:Combination}.
In Fig.~\ref{fig:Example:PriorSNR}, the estimated \emph{a priori} \acp{SNR}~$\sPriorSNRI$ obtained from the non-\ac{MLSE}-based enhancement scheme shows a fine structure which is similar to the speech structure visible in Fig.~\ref{fig:Example:SpectrogramNoisy}.
Contrarily, the structure of the \emph{a priori} \acp{SNR}~$\sPriorSNRI$ estimated by the \ac{MLSE}-based enhancement schemes is very coarse and reveals no or only little of the harmonic fine structure shown in Fig.~\ref{fig:Example:SpectrogramNoisy}.
Using these envelope models for the speech component leads to an overestimation of the \emph{a priori} \acp{SNR}~$\sPriorSNRI$ between spectral harmonics.

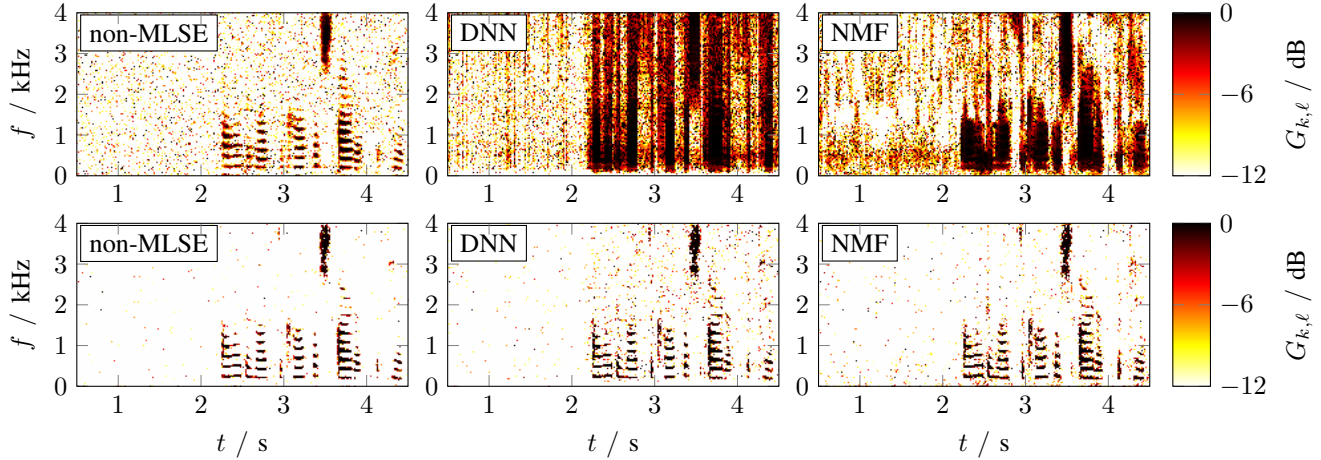
\begin{figure*}[tb]
    \centering
    \begin{tikzpicture}
    \begin{groupplot}[
            group style={%
                group size=3 by 2,
                xlabels at=edge bottom,
                ylabels at=edge left,
                vertical sep=4ex,
                horizontal sep=1.5em,
            },
            width=0.33\textwidth,
            height=3.75cm,
            specgram_axis,
            point meta min=-12,
            point meta max=0,
            colorbar style={ylabel={$\sGainI~/~\text{dB}$}, ytick={-12,-6,0}}
        ]
        \edef\tmp{}
        \newcounter{gfxcount}{0}
        \edef\colorbar{false}
        \pgfplotsforeachungrouped \method/\displayname in {generic/non-MLSE, nn/DNN, nmf/NMF} {%
            \ifnum\value{gfxcount} > 1
                \edef\colorbar{true}
            \fi
            \eappto\tmp{\noexpand\nextgroupplot[colorbar=\colorbar]}
            \eappto\tmp{\noexpand\getImage{Data/example/gain_\method_CBetaOrder_1.00.tsv}}
            \eappto\tmp{\noexpand\node[anchor=north west, fill=white, rectangle, draw] at(rel axis cs: 0.01, 0.99) {\displayname};}
            \stepcounter{gfxcount}
        }
        \setcounter{gfxcount}{0}
        \edef\colorbar{false}
        \pgfplotsforeachungrouped \method/\displayname in {generic/non-MLSE, nn/DNN, nmf/NMF} {%
            \ifnum\value{gfxcount} > 1
                \edef\colorbar{true}
            \fi
            \eappto\tmp{\noexpand\nextgroupplot[colorbar=\colorbar]}
            \eappto\tmp{\noexpand\getImage{Data/example/gain_\method_CBetaOrder_0.20.tsv}}
            \eappto\tmp{\noexpand\node[anchor=north west, fill=white, rectangle, draw] at(rel axis cs: 0.01, 0.99) {\displayname};}
            \stepcounter{gfxcount}
        }
        \tmp
    \end{groupplot}
\end{tikzpicture}\\
    \caption{\label{fig:Example:GainGauss}Gain applied to the noisy input coefficients~$\sNoisyFreqI$ by \ac{MOSIE}~\cite{breithaupt_parameterized_2008} for different \ac{MLSE}-based enhancement schemes for the excerpt shown in Fig.~\ref{fig:Example:SpectrogramNoisy}.
    In the upper row, $\sShape = 1$ and $\sCompression = 0.001$ which approximates the Gaussian \ac{LSA} proposed in~\cite{ephraim_speech_1985} as shown in Table~\ref{tab:RelatedEstimators}.
    In the lower, $\sShape = 0.2$ and $\sCompression = 0.001$ is used which corresponds to a super-Gaussian~\ac{LSA}.
    Here, $f$ denotes frequency and $t$ time.
    }
\end{figure*}

Next, the gain as defined in~\eqref{eq:GainFunction} is considered.
For this example, we use \ac{MOSIE}~\cite{breithaupt_parameterized_2008} with two different parameter setups.
First, a setup is used where the clean speech coefficients~$\sSpeechFreqI$ are assumed to follow a complex circular-symmetric Gaussian distribution.
For this, the parameters of \ac{MOSIE}~\cite{breithaupt_parameterized_2008} are set to $\sShape = 1$ and $\sCompression = 0.001$, which approximates the Gaussian \ac{LSA}~\cite{ephraim_speech_1985}.
For the second setup, the shape is reduced to $\sShape = 0.2$, i.e., a super-Gaussian \ac{LSA} is employed.
To limit speech distortions, the gain is limited such that attenuations larger than 12~dB are prevented.
This limit is applied throughout the paper if not stated otherwise.
The applied gains for the Gaussian and super-Gaussian case are shown in Fig.~\ref{fig:Example:GainGauss}.

The upper row in Fig.~\ref{fig:Example:GainGauss} shows that the overestimations of the \emph{a priori} \ac{SNR}~$\sPriorSNRI$, e.g., between spectral harmonics, result in a poor suppression for the \ac{MLSE}-based enhancement schemes when using a Gaussian estimator $(\sShape = 1)$.
The non-\ac{MLSE}-based enhancement scheme is, however, not affected and achieves high suppression values between harmonics.
As discussed in Section~\ref{sec:Analysis}, this behavior can be explained from Fig.~\ref{fig:MOSIEPrior}.
For $\sShape = 1$, the attenuation is mainly controlled by the \emph{a priori} \ac{SNR}~$\sPriorSNRI$ where lower \emph{a priori} \acp{SNR}~$\sPriorSNRI$ lead to higher suppression values.
From this it follows that an overestimation of $\sPriorSNRI$ results in lower attenuations as observed for the \ac{MLSE}-based enhancement schemes.
As a consequence, using Gaussian clean speech estimators (see Table~\ref{tab:RelatedEstimators}) for \ac{MLSE}-based enhancement schemes results in audible artifacts.

Interestingly, the lower row in Fig.~\ref{fig:Example:GainGauss} shows that the issues observed for $\sShape = 1$ can be reduced if a super-Gaussian estimator $(\sShape < 1 )$ is employed.
In contrast to Fig.~\ref{fig:Example:GainGauss}, noise is suppressed also between harmonics.
Further, also higher attenuations are applied to the noise only segments.
Considering Fig.~\ref{fig:MOSIEPost} and Fig.~\ref{fig:MOSIEPrior}, the behavior can be explained by the fact that lower shape values cause more suppression for low \emph{a posteriori} \acp{SNR}~$\sPostSNRI$.
Hence, our key conclusion is that using super-Gaussian clean speech estimators, the background noise can be suppressed also when \ac{MLSE}-based approaches are employed.

\section{Instrumental Evaluation}
\label{sec:Evaluation}

We evaluate the performance of the different speech estimators using instrumental measures such as \ac{PESQ} improvement scores~\cite{noauthor_p.862:_2001} and \ac{SegSNR} improvements~\cite{lotter_speech_2005, gerkmann_unbiased_2012}.
The improvements are based on the noisy signal, i.e., they are computed as the difference between the raw scores of the enhanced signal and the noisy signal.
Additionally, the \ac{SegSSNR} and the \ac{SegNR}~\cite{lotter_speech_2005} are employed to quantify the speech distortions and noise suppression, respectively.
Higher values for the \ac{SegSSNR} indicate less speech distortion and higher values for the \ac{SegNR} indicates more noise reduction.

For this evaluation, we use 128~sentences from the TIMIT core set.
Again, it is ensured that the amount of audio material is balanced between genders.
The clean speech signals are artificially corrupted by the same noise types used for training the \ac{NMF}-based enhancement scheme.
The \acp{SNR} are ranging from -5~dB to 20~dB in 5~dB steps.
For each sentence, the segment of the noise signal where the speech signals are embedded in is randomly chosen.
The instrumental measures are only evaluated after a two second initialization period to avoid initialization artifacts that may bias the results.
Similarly, also the \acp{SNR} used for the artificial mixing are determined based on the signal powers in speech presence.
Further, the noise segments that were used for training the \ac{NMF}-based enhancement scheme are excluded in the evaluation for all enhancement schemes, i.e., also for the non-\ac{MLSE}-based and the \ac{DNN} based enhancement schemes.
This is done to make the enhancement schemes more easily comparable.

\subsection{Performance Impact of \acs{MOSIE}'s Parameters}

In this section, we analyze how the choice of the shape and the compression parameter influences the performance of clean speech estimators if used for the \ac{MLSE}-based enhancement schemes.

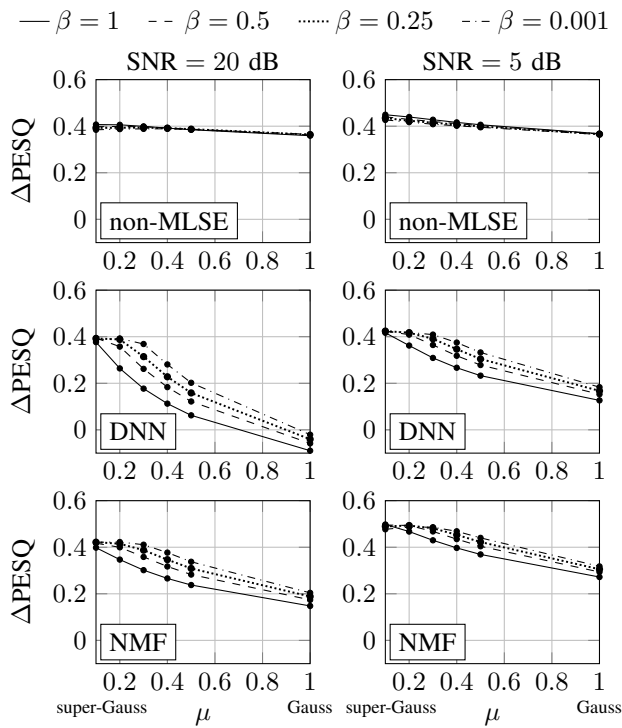
\begin{figure}[tb]
    \centering
    \begin{tikzpicture}
    \begin{customlegend}[
            legend entries={$\sCompression = 1$, $\sCompression = 0.5$, $\sCompression = 0.25$, $\sCompression = 0.001$},
            legend columns=-1,
            legend style={draw = none, inner sep = 2pt, /tikz/every even column/.append style={column sep=0.3cm}},
        ]
        \addlegendimage{comp1.0000};
        \addlegendimage{comp0.5000};
        \addlegendimage{comp0.2500};
        \addlegendimage{comp0.0010};
    \end{customlegend}
\end{tikzpicture}\\
    \begin{tikzpicture}
    \begin{groupplot}[
            group style={%
                group name=ShapeBetaOrder,
                group size=2 by 3,
                xlabels at=edge bottom,
                ylabels at=edge left,
                vertical sep=4ex,
            },
            xmin=0.1,
            xmax=1.0,
            ymin=-0.1,
            ymax=0.6,
            grid,
            width=0.5\linewidth,
            height=3.75cm,
            xlabel={$\sShape$},
            ylabel={$\Delta$\acs{PESQ}},
        ]

        \edef\tmp{}
        \pgfplotsforeachungrouped \method/\methodhuman in {generic/non-MLSE, dnn_poslin/DNN, nmf/NMF} {%
            \pgfplotsforeachungrouped \snr in {20, 5} {%
                \eappto\tmp{\noexpand\nextgroupplot}
                \pgfplotsforeachungrouped \comp in {1.0000, 0.5000, 0.2500, 0.0010} {%
                    \eappto\tmp{\noexpand\addplot[comp\comp, mymarks] table[x=shape, y=pesq_imp] {Data/shape/shape_\method_betaorder_\snr_\comp.tsv};}
                }
                \eappto\tmp{\noexpand\node[fill=white, draw, anchor=south west] at (rel axis cs:0.02, 0.02) {\methodhuman};}
            }
        }
        \tmp
    \end{groupplot}
    \node[anchor=south] at (ShapeBetaOrder c1r1.north) {$\text{SNR} = 20~\text{dB}$};
    \node[anchor=south] at (ShapeBetaOrder c2r1.north) {$\text{SNR} = 5~\text{dB}$};

    \node[xshift=0.5ex, yshift=-1.5ex, label={[font=\scriptsize]below:super-Gauss}] at (ShapeBetaOrder c1r3.south west) {};
    \node[yshift=-1.5ex, label={[font=\scriptsize]below:Gauss}] at (ShapeBetaOrder c1r3.south east) {};
    \node[xshift=0.5ex, yshift=-1.5ex, label={[font=\scriptsize]below:super-Gauss}] at (ShapeBetaOrder c2r3.south west) {};
    \node[yshift=-1.5ex, label={[font=\scriptsize]below:Gauss}] at (ShapeBetaOrder c2r3.south east) {};
\end{tikzpicture}\\
    \caption{\label{fig:PESQBetaOrder}\acs{PESQ} improvement scores of \ac{MOSIE}~\cite{breithaupt_parameterized_2008} for all considered enhancement schemes in dependence of the shape~$\sShape$ and compression~$\sCompression$.
        For relations to other clean speech estimators, see Table~\ref{tab:RelatedEstimators}.
    }
\end{figure}

Fig.~\ref{fig:PESQBetaOrder} shows the \ac{PESQ} improvement scores for \ac{MOSIE}~\cite{breithaupt_parameterized_2008} as a function of the shape parameter~$\sShape$ and the compression parameter~$\sCompression$.
The graphs depict the average over all considered noise types and speech files for two different input \acp{SNR}.
For the non-\ac{MLSE}-based enhancement scheme, increasing super-Gaussianity $(\sShape < 1)$ and compression $(\sCompression < 1)$ slightly improve the predicted speech quality by \ac{PESQ}.
However, the key message is that for the \ac{MLSE}-based enhancement schemes, increasing super-Gaussianity ($\sShape < 1$) and compression ($\sCompression < 1$) improve the signal quality predicted by \ac{PESQ} considerably stronger.

\subsection{Comparison with Common Enhancement Schemes}

In this final part of the evaluation section, we compare the super-Gaussian estimators, i.e., \ac{MOSIE}~\cite{breithaupt_parameterized_2008} to Gaussian approaches.
To demonstrate that super-Gaussian estimators considerably improve the performance of \ac{MLSE}-based methods, we use the following two parameter settings
for \ac{MOSIE}~\cite{breithaupt_parameterized_2008}: $\sCompression = 0.001, \sShape = 0.2$ and $\sCompression = 1, \sShape = 0.2$.
The parameters are chosen as a compromise such that all \ac{MLSE}-based enhancement schemes yield satisfying results.

\begin{figure*}[tb]
    \centering
    \begin{tikzpicture}
    \begin{customlegend}[
            legend entries={
                {LSA ($\sShape = 1$, $\sCompression = 0.001$)},
                {STSA ($\sShape = 1$, $\sCompression = 1$)},
                {MOSIE ($\sShape = 0.2$, $\sCompression = 0.001$)},
                {MOSIE ($\sShape = 0.2$, $\sCompression = 1$)},
            },
            legend columns=-1,
            legend style={draw=none, inner sep=2pt, /tikz/every even column/.append style={column sep=0.3cm}},
        ]
        \addlegendimage{LSA};
        \addlegendimage{STSA};
        \addlegendimage{MOSIE};
        \addlegendimage{MOSIESTSA};
    \end{customlegend}
\end{tikzpicture}\\[0.1ex]
    \begin{tikzpicture}
    \begin{groupplot}[
            ax seg_nr/.style={%
                ymin=0, ymax=14,
            },
            ax seg_ssnr/.style={%
                ymin=2.5, ymax=25,
            },
            ax segsnr_imp/.style={%
                ymin=0, ymax=11,
            },
            ax pesq_imp/.style={%
                ymin=-0.1,
                ymax=0.6,
            },
            group style={%
                group size=3 by 4,
                xlabels at=edge bottom,
                ylabels at=edge left,
                vertical sep=4ex,
                group name=Comparison Eval,
            },
            width=0.35\textwidth,
            height=3.5cm,
            grid,
            xlabel={SNR / dB},
            xmin=-5,
            xmax=20,
        ]

        \edef\tmp{}
        \pgfplotsforeachungrouped \measure/\measurehuman in {pesq_imp/{$\Delta$PESQ}, segsnr_imp/{$\Delta$SegSNR / dB}, seg_ssnr/{SegSSNR / dB}, seg_nr/{SegNR / dB}} {%
            \pgfplotsforeachungrouped \method in {generic, dnn_poslin, nmf} {%
                \eappto\tmp{\noexpand\nextgroupplot[ylabel={\measurehuman}, ax \measure]}
                \eappto\tmp{\noexpand\addplot[LSA] table[x=snr, y=\measure] {Data/results/results_\method_betaorder_1.00_0.0010.tsv};}
                \eappto\tmp{\noexpand\addplot[STSA] table[x=snr, y=\measure] {Data/results/results_\method_betaorder_1.00_1.0000.tsv};}
                \eappto\tmp{\noexpand\addplot[MOSIE] table[x=snr, y=\measure] {Data/results/results_\method_betaorder_0.20_0.0010.tsv};}
                \eappto\tmp{\noexpand\addplot[MOSIESTSA] table[x=snr, y=\measure] {Data/results/results_\method_betaorder_0.20_1.0000.tsv};}
                \edef\measurehuman{}
            }
        }
        \tmp
    \end{groupplot}
    \node[anchor=south] at(Comparison Eval c1r1.north) {non-MLSE};
    \node[anchor=south] at(Comparison Eval c2r1.north) {DNN};
    \node[anchor=south] at(Comparison Eval c3r1.north) {NMF};
\end{tikzpicture}\\
    \caption{\label{fig:LargeComparison}\acs{PESQ} improvement scores and segmental \ac*{SNR} measures for different clean speech estimators employed in the non-\ac{MLSE}-based, the \ac{DNN} based, and the \ac{NMF} based enhancement scheme.
        While \acs{LSA} and \acs{STSA} employ Gaussian speech priors, \ac{MOSIE} ($\sShape = 0.2, \sCompression = 0.001$) and \ac{MOSIE} ($\sShape = 0.2, \sCompression = 1$) represent modern super-Gaussian speech estimators (see Table~\ref{tab:RelatedEstimators}).
    }
\end{figure*}
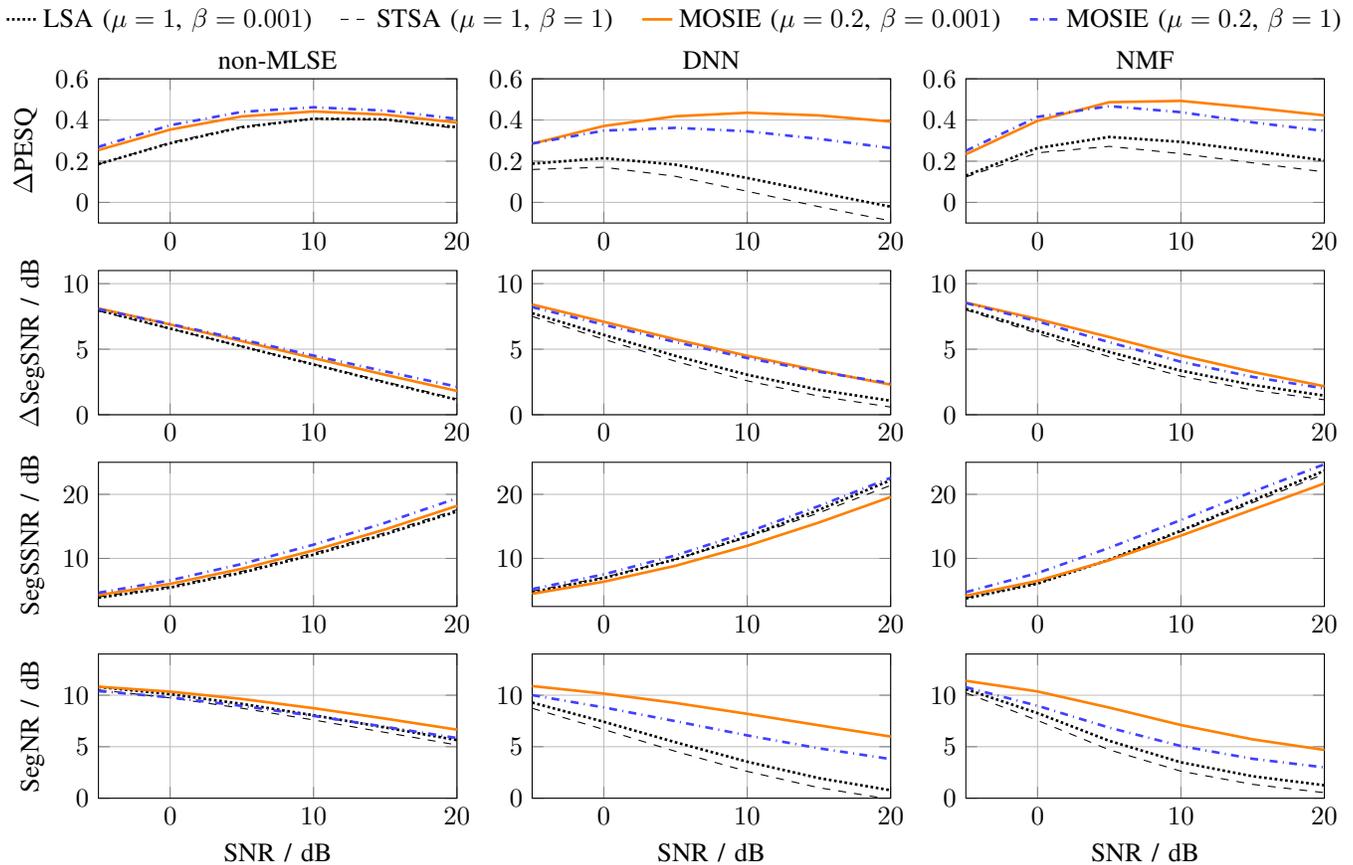

Fig.~\ref{fig:LargeComparison} shows \ac{PESQ} improvement scores and segmental \ac{SNR} measures for the considered enhancement schemes.
The results again show that for the non-\ac{MLSE}-based enhancement scheme, a super-Gaussian estimator only slightly improves the performance.
Contrarily, the super-Gaussian setup for \ac{MOSIE}~\cite{breithaupt_parameterized_2008} performs considerably better than the Gaussian clean speech estimator, i.e., the Gaussian \ac{STSA}~\cite{ephraim_speech_1984} and the Gaussian \ac{LSA}~\cite{ephraim_speech_1985}, if the \ac{MLSE}-based estimators are considered.
As shown in Section~\ref{sec:Analysis}, the suppression capability of the Gaussian approaches is mainly controlled by the \emph{a priori} \ac{SNR} resulting in low suppressions between harmonics for the \ac{MLSE}-based enhancement schemes where the \emph{a priori} \ac{SNR} is overestimated.
Here, this is reflected by the low segmental noise reduction values observed for the \ac{DNN}-based and the \ac{NMF}-based approach if the Gaussian \ac{STSA}~\cite{ephraim_speech_1984} or the Gaussian \ac{LSA}~\cite{ephraim_speech_1985} are employed.
However, for the super-Gaussian estimators \ac{MOSIE} $(\sShape = 0.2, \sCompression = 0.001)$ and \ac{MOSIE} $(\sShape = 0.2, \sCompression = 1)$ the noise reduction is strongly increased and the residual noise, e.g., the noise between harmonics, is reduced.
This comes with a slight increase in speech distortion for \ac{MOSIE} $(\sShape = 0.2, \sCompression = 0.001)$ as visible in a decrease in \ac{SegSSNR}.
For \ac{MOSIE} ($\sShape = 0.2, \sCompression = 1)$, the \ac{SegSSNR} remains unchanged or is even slightly increased.
Overall, the behavior of the super-Gaussian estimators helps to improve the quality predicted by \ac{PESQ} and to improve the \ac{SegSNR}.

\section{Subjective Evaluation}
\label{sec:SubjectiveEvaluation}

As the results of instrumental measures cannot perfectly represent the impressions of human listeners, we verify the results using a subjective listening test.
For this, we employ a \ac{MUSHRA}~\cite{noauthor_bs.1534-3:_2015}.
In the experiment, two different acoustic scenarios are tested: traffic noise and babble noise both at an \ac{SNR} of 5~dB.
For both acoustic scenarios, an utterance of a male and a female speaker taken from the TIMIT test set are used.
These signals are processed by the non-\ac{MLSE}-based enhancement scheme, the \ac{DNN}-based enhancement scheme, and the \ac{NMF}-based enhancement schemes.
For all enhancement schemes, a Gaussian \ac{STSA} ($\sShape = 1, \sCompression = 1$) and a super-Gaussian \ac{STSA} ($\sShape = 0.2, \sCompression = 1$) are compared (see Table~\ref{tab:RelatedEstimators}).
Even though \ac{MOSIE} with $\sShape = 0.2$ and $\sCompression = 0.001$ achieves the highest scores in most instrumental measures, we use \ac{MOSIE} with $\sCompression = 1$ in the subjective listening test as this configuration produces less musical artifacts.

In each trial, four signals are presented to the listeners: the noisy signals processed by the Gaussian and the super-Gaussian estimator, an anchor, and a hidden reference.
The trials are repeated over all combinations of acoustic conditions, speakers and enhancement schemes.
The reference signal is a noisy signal with an \ac{SNR} 17~dB.
Finally, for the anchor, the clean speech utterance is filtered using a low-pass filter at a cutoff frequency of 4~kHz and mixed at an \ac{SNR} of $-5~\text{dB}$.
This signal is processed using a non-\ac{MLSE}-based enhancement scheme where the noise \ac{PSD} is estimated using~\cite{gerkmann_noise_2011} and the speech \ac{PSD} is obtained using the decision-directed approach~\cite{ephraim_speech_1984} with a smoothing constant set to $0.9$.
A Wiener filter with a minimum gain of $-20~\text{dB}$ is employed to obtain the anchor.
The sound examples used in the experiment are also available at \url{https://uhh.de/inf-sp-tasl2018a}.

A total of 13 subjects have participated in the \ac{MUSHRA}.
The test took place in a quiet office and the subjects listened to diotic signals played back through headphones (Beyerdynamic DT-770 Pro 250~Ohm) through a RME Fireface UFX+ sound card.
The test was conducted in two phases.
In the first phase, the subjects were asked to listen to a subset of the files used in test such that they can familiarize themselves with the different signals.
During this training phase, the listeners were also asked to set the level of the headphones to a comfortable level.
In the second phase, the listener's task was to judge the overall quality of the signals on a scale ranging from 0 to 100, where 0 was labeled with “bad” and 100 with “excellent”.
The order of presentations of algorithms and conditions were randomized between all subjects.

\begin{figure}[tb]
    \centering
    \hspace*{2em}\begin{tikzpicture}
    \begin{customlegend}[
            legend entries={reference, anchor, Gauss, super-Gauss},
            legend cell align={left},
            legend columns=-1,
            legend style={draw=none, inner sep=1pt},
        ]
        \addlegendimage{area legend, fill=violet!0};
        \addlegendimage{area legend, fill=violet!33};
        \addlegendimage{area legend, fill=violet!66};
        \addlegendimage{area legend, fill=violet!99};
    \end{customlegend}
\end{tikzpicture}\\
    \begin{tikzpicture}[]
    \begin{axis}[%
        scale only axis,
        width=0.75\linewidth,
        height=3cm,
        ylabel={MUSHRA score},
        grid,
        boxplot/draw direction=y,
        boxplot/every median/.style={very thick},
        boxplot/every box/.append style={fill=white},
        boxplot/box extend=0.1,
        xtick={1,2,3},
        xmin=0.5,
        xmax=3.5,
        xticklabels={DNN, NMF, non-MLSE},
        title style={anchor=south, at={(0.5, 0.925)}},
        ]
        \edef\tmp{}
        \pgfmathsetmacro{\i}{0}
        \pgfplotsforeachungrouped \algo/\algohuman in {nn/DNN based enhacement scheme,nmf/NMF based enhancement scheme,gen/non-MLSE based enhancement scheme} {
            \pgfmathtruncatemacro{\i}{\i + 1}
            \pgfmathsetmacro{\j}{0}
            \pgfplotsforeachungrouped \stimulus in {reference, anchor, gauss, supergauss} {
                \pgfmathtruncatemacro{\q}{33*\j}
                \eappto\tmp{\noexpand\addplot[boxplot, boxplot/every box/.style={fill=violet!\q}, xshift=10*(\j-1.5), boxplot/draw position=\i, mark=x] table[y=\stimulus] {Data/boxdata/\algo_boxdata.tsv};}
                \pgfmathsetmacro{\j}{\j + 1}
            }
        }
        \tmp
    \end{axis}
\end{tikzpicture}\\
    \caption{\label{fig:ListeningExperiment}Box plot of the subjective ratings for different enhancement schemes.}
\end{figure}
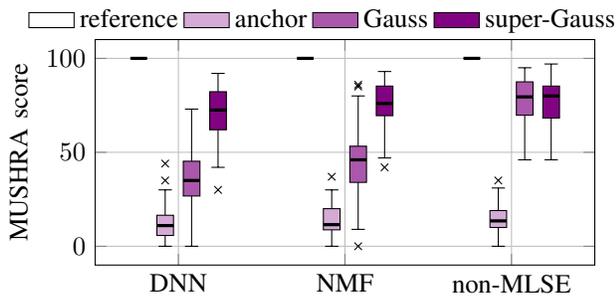

The obtained \ac{MUSHRA} scores are summarized in Fig.~\ref{fig:ListeningExperiment} using box plots.
The upper and the lower edge of the box show the upper and lower quartile while the bar within the box is the median.
The upper whisker reaches to the largest data point that is smaller than the upper quartile plus 1.5 times the interquartile range.
The lower whisker is defined analogously.
The crosses denote outliers that do not fall in the range spanned by both whiskers.
For each box plot, the results of all acoustic conditions and speakers are pooled, which yields 52 data points.
The result show that all participants were able to detect the hidden reference, which had to be rated with 100, and that the anchor was consistently given the lowest scores.
Further, the results clearly confirm that for the \ac{DNN} and the \ac{NMF} based enhancement scheme, the sound quality of the super-Gaussian estimator is considered better than the Gaussian estimator.
For the non-\ac{MLSE}-based estimator, however, the \ac{MUSHRA} scores of the Gaussian and the super-Gaussian estimator are nearly the same.

Finally, a brief statistical analysis of the results confirms that the differences in \ac{MUSHRA} scores between the Gaussian and super-Gaussian estimators are statistically significant for the \ac{MLSE}-based enhancement schemes.
For the used statistical tests, a significance level of $\sSigLevel = 0.05$ is employed.
We apply a Wilcoxon signed-rank test to test for the difference in medians between the \ac{MUSHRA} scores of the Gaussian and super-Gaussian estimators.
This test is employed as the Shapiro-Wilk test indicates that the data is not Gaussian distributed for all conditions.
The different enhancement schemes, i.e., the \ac{MLSE}-based approaches and the non-\ac{MLSE}-based approach, are treated separately.
Considering the difference between the Gaussian and super-Gaussian clean speech estimators for the \ac{MLSE}-based approaches, the differences are significant  in both cases (\ac{DNN}: $p < 0.001$, \ac{NMF}: $p < 0.001$).
Comparing the estimators for the non-\ac{MLSE}-based algorithm reveals no significant difference ($p = 0.55$).
Hence, the subjective listening tests confirm the previously obtained results of the instrumental measures.

\section{Conclusions}
\label{sec:Conclusions}

In this paper, super-Gaussian clean speech estimators have been analyzed in the context of machine-learning based speech enhancement approaches that employ spectral envelope models.
We refer to these approaches as \ac{MLSE}.
In the analysis part, we showed that the usage of envelope models results in an overestimation of the \emph{a priori} \ac{SNR}, e.g., between speech spectral harmonics.
As a consequence, using Gaussian estimators, noise between harmonic structures cannot be reduced such that residual noises remain after the enhancement.
However, in this paper, we show that employing super-Gaussian clean speech estimators, such as \ac{MOSIE}~\cite{breithaupt_parameterized_2008}, leads to a reduction of the undesired residual noise.
This interesting result stems from the higher attenuation that is applied by the super-Gaussian estimators if the \emph{a posteriori} \acp{SNR} are low.
This allows the estimators to compensate for the overestimated \emph{a priori} \acp{SNR} without any further post-processing steps.
As a consequence, we showed via theoretical analysis and experimental evaluation that for \ac{MLSE}-based enhancement schemes, super-Gaussian estimators have a much larger effect on improving the enhancement performance than for classic non-\ac{MLSE}-based enhancement schemes.
Sound examples of the considered algorithms are given at \url{https://uhh.de/inf-sp-tasl2018a}.

\ifCLASSOPTIONcaptionsoff
  \newpage
\fi

\bibliographystyle{IEEEtran}
\bibliography{Bibliography/Journal2016}

\begin{IEEEbiography}[{\includegraphics[width=1in,height=1.25in,clip,keepaspectratio]{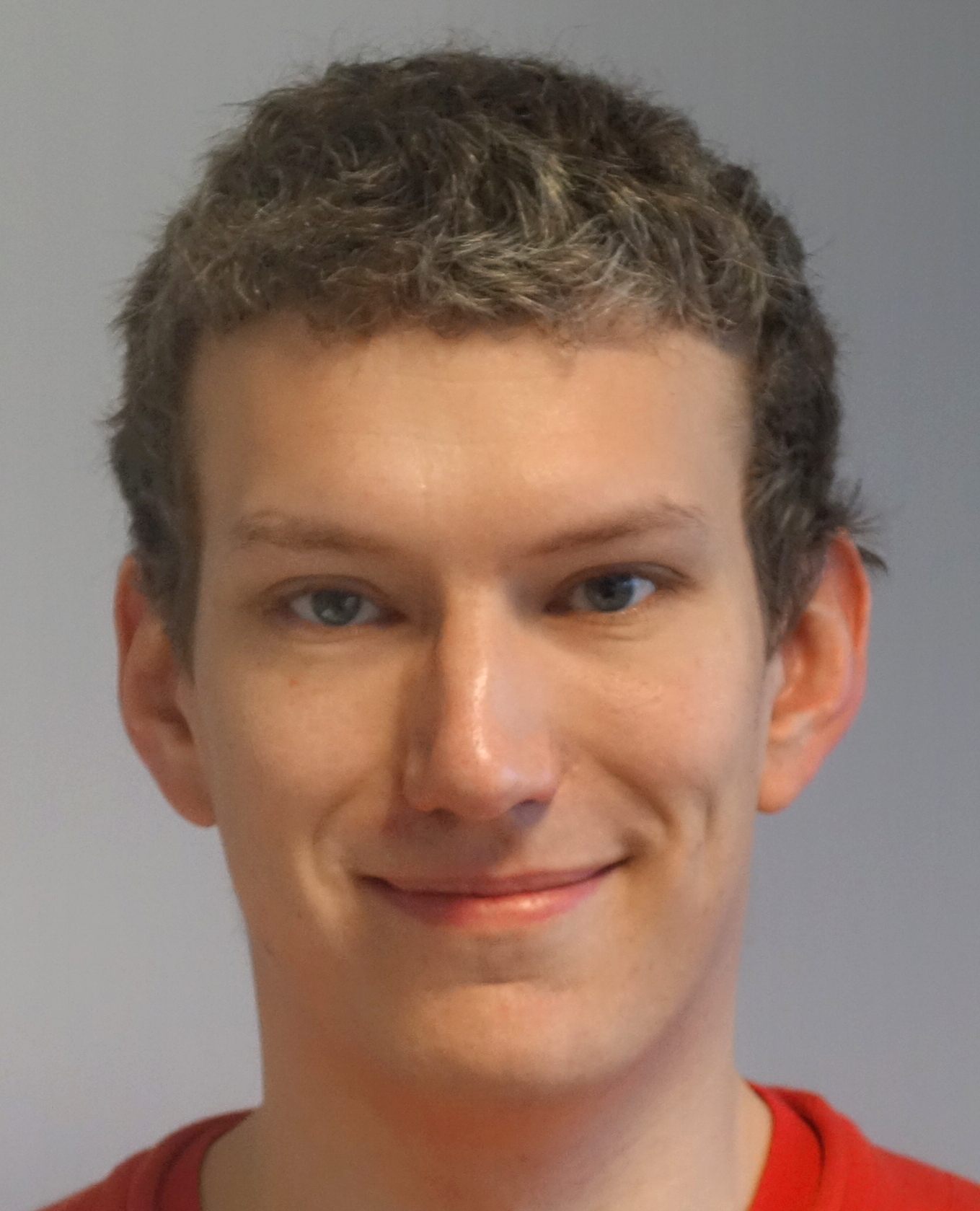}}]{Robert Rehr}
     received his B.Eng.\ degree from the Jade Hochschule, Oldenburg, Germany, in 2011, and the M.Sc.\ degree from the Universität Oldenburg, Oldenburg, Germany, in 2013, both in Hearing Technology and Audiology.
    From 2013 to 2016, he was with the Speech Signal Processing Group at the Universität Oldenburg, Oldenburg, Germany.
    Since 2017 he has been with the Signal Processing Group at the Universität Hamburg, Hamburg, Germany.
    R. Rehr is currently pursuing the Ph.D. degree.
    His research focuses on speech enhancement using machine-learning based methods.
\end{IEEEbiography}

\begin{IEEEbiography}[{\includegraphics[width=1in,height=1.25in,clip,keepaspectratio]{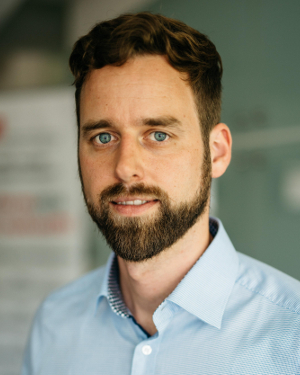}}]{Timo Gerkmann}
     received the Dipl.-Ing.\ and Dr.-Ing.\ degrees in electrical engineering and information sciences from the Ruhr-Universität Bochum, Bochum, Germany, in 2004 and 2010 respectively.
    In 2005, he spent six months with Siemens Corporate Research, Princeton, NJ, USA. From 2010 to 2011, he was a Postdoctoral Researcher in the Sound and Image Processing Laboratory, Royal Institute of Technology (KTH), Stockholm, Sweden.
    From 2011 to 2015, he was a Professor of speech signal processing with the Universität Oldenburg, Oldenburg, Germany.
    From 2015 to 2016, he was the Principal Scientist in Audio \& Acoustics, Technicolor Research \& Innovation, Hanover, Germany.
    Since 2016, he has been a Professor of signal processing with the University of Hamburg, Hamburg, Germany.
    His research interests include digital signal processing algorithms for speech and audio applied to communication devices, hearing instruments, audio–visual media, and human–machine interfaces.
    He is member of the IEEE Signal Processing Society Technical Committee on Audio and Acoustic Signal Processing.
\end{IEEEbiography}

\vfill

\end{document}